\documentclass[12pt]{article}  
\usepackage{epsfig}  
\def\tozero#1{\mathrel{\mathop{\sim}\limits_{\scriptscriptstyle
{#1\rightarrow0 }}}}
\def\bea{\begin{eqnarray}}
\def\eea{\end{eqnarray}}
\def\beq{\begin{equation}}  
\def\eeq{\end{equation}}

\def\as{\alpha_s}
\def\az{\alpha_0}
\relax  
 
\def\zalpha{Z^{(\as)}(\as(\mu^2),\epsilon)} 
\def\zalphaQ{Z^{(\as)}(\as(Q^2),\epsilon)} 
\def\zc{Z^{(C)}(N,\as(\mu^2),\epsilon)} 

\def\eg{{\it e.g.}}  
\def\ie{{\it i.e.}}  
\def\MS{\hbox{$\overline{\rm MS}$}}  
\def    \hepph  #1 {{\tt hep-ph/#1}}  
\def    \hepex  #1 {{\tt hep-ex/#1}}  
  
\newcommand{\sect}[1]{\setcounter{equation}{0}\section{#1}}

\newsavebox\tmpfig  
  
\begin{document}  
  
\pagestyle{empty}  
  
\begin{flushright}   
  
{\tt hep-ph/0209154}\\RM3-TH/02-02\\ GeF/TH/11-02\\
\end{flushright}   
  
\begin{center}   
\vspace*{0.5cm}  
{\Large \bf Renormalization group approach \\ to soft gluon resummation} \\   
\vspace*{1.5cm}   
Stefano Forte$^{a}$  
and   
Giovanni Ridolfi$^{b}$\\  
\vspace{0.6cm}  {\it
{}$^a$INFN, Sezione di Roma Tre\\  
Via della Vasca Navale 84, I-00146 Roma, Italy\\ \medskip
{}$^b$INFN, Sezione di Genova,\\  
Via Dodecaneso 33, I-16146 Genova, Italy}\\  
\vspace*{1.5cm}  
  
{\bf Abstract}  
  
\end{center}  
  
\noindent  
We present a simple proof of the all-order exponentiation of soft
logarithmic corrections to hard processes in perturbative QCD. Our argument
is based on proving that all large logs in the soft limit
can be expressed in terms
of a single dimensionful variable, and then using the renormalization
group to resum them. Beyond the next-to-leading log level, our result
is somewhat less predictive than previous all-order resummation
formulae, but it does not rely on non-standard factorization, and it is thus
possibly more general. We
use our result to settle issues of convergence of the resummed
series, we  discuss scheme dependence at the resummed level, 
and we provide explicit resummed expressions in various
factorization schemes.
\\  
\vspace*{1cm}  

\vfill  
\noindent  
  
\begin{flushleft} September 2002 \end{flushleft}   
\eject   
  
\setcounter{page}{1} \pagestyle{plain}  
  
\sect{The structure of soft resummation}
\label{intro}

The resummation of large logarithms related to soft radiation
 near the boundary of phase space is a classic subject, which
stems from very early results in quantum
electrodynamics~\cite{sudakov}, and is of great phenomenological and
theoretical interest for QCD at colliders~\cite{resrev}.  The
exponentiation of soft logs related to gluon emission
has been proven in QCD to leading
order~\cite{loexp}, and to next-to-leading order (NLO) through an
explicit analysis of the relevant Feynman graphs, either using
eikonal~\cite{cnt} or factorization~\cite{sterman} techniques, which
lead to equivalent~\cite{equiv} resummation formulae. These
resummation formuale were subsequently used~\cite{paolo} to obtain
resummed results for physical cross sections.

More recently~\cite{contopa}, it was shown that a large class of
resummation formulae can be derived in a unified way and to all orders
by assuming the validity of a suitable factorization~\cite{fact}, and
then applying the renormalization group to it. The price to pay for
this generality is that one has to rely on a rather nontrivial
two-scale factorization, which goes beyond the standard factorization
of collinear large logs. Also, the resummed result obtained in this
way looks quite formal: in particular, its equivalence with
resummation formulae of the form of Refs.~\cite{cnt,sterman} is only
established~\cite{contopa} to next-to-leading order.

In this paper, we present a simple argument for the all-order
resummation of soft logs. Our argument is based on the following
steps. First, we establish to all orders the relation between the
resummation of large logs of the kinematic variable $1-x$, and large
logs of its Mellin conjugate variable $N$, up to power-suppressed
terms in the soft limit $x\to 1$, \ie\ $N\to\infty$. This relation
generalizes to all orders the relation between the next-to-leading
resummations of Refs.~\cite{cnt} and \cite{sterman}, which can be
respectively viewed as a resummation of $\ln(1-x)$ contributions to
the splitting function, and a resummation of $\ln\frac{1}{N}$
contributions to the anomalous dimension.

Next, we prove, that the large logarithms
of $(1-x)$ all stem from logs of a dimensionful variable
$Q^2(1-x)^a$, where $a$ is an integer which characterizes the physical
process. This origin of this
result is essentially kinematic. It is  related to the
structure of the phase space for tree-level $n$-particle emission, 
which turns out not to be
modified by loop corrections.  Finally,
we resum the logs of this latter dimensionful variable using standard
renormalization-group techniques. This leads to a resummed result
which 
generalizes to all orders the next-to-leading log resummation of
Refs.~\cite{cnt,sterman}
in a way which is somewhat less predictive than the
all-order result of Ref.~\cite{contopa}. Indeed, we also conclude,
like Ref.~\cite{contopa}, that logs of $N$ exponentiate to all
orders. However, using the resummation formula of Ref.~\cite{contopa}
the N$^{k-1}$LO resummed result is determined by a fixed-order N$^k$LO
calculation, whereas using our resummation the N$^{k-1}$LO resummed result
is determined by a fixed-order N$^{\frac{k(k+1)}{2}-1}$LO
computation. 
On the other
hand, our result does not require two-scale factorization, and it is
thus easier to derive, and possibly more general. 

We then exploit our relation between the resummation of logs of $N$
and $1-x$ to study the relation between various resummed results.
First, given that the resummed result can be equivalently viewed as a
$\ln(1-x)$ resummation of the splitting function or a $\ln\frac{1}{N}$
resummation of the anomalous dimension, we discuss the possibility of
further casting it as a $\ln(1-x)$ resummation of the physical cross
section. We show that, whereas this is formally possible, the ensuing
result is meaningless since it is the sum of a factorially
divergent~\cite{paolo} series. We trace the origin of this divergence,
and we show that it is related to subleading logs, and thus in
particular unrelated to power corrections.  We then turn to the
derivation of explicit expressions for the resummed cross section in a
variety of factorization schemes, and discuss the wider freedom in
choice of factorization which is available at the resummed level.

\sect{Next-to-leading log resummation}  
\label{NLLres}

In this section, we review the well-known resummation formulae of
Refs.~\cite{cnt,sterman}, and recast them in a form which is best
suited to our subsequent derivation and generalization.

We consider the total cross section for a partonic hard process in the
vicinity of the kinematic boundary for the production of the final
state. We assume that the kinematics of the
process is specified by a hard scale $Q^2$, and a dimensionless
variable $x$, defined
in such a way that $0\le x\le 1$, and $x=1$ is
the kinematic boundary, \ie\ the limit in which there is no phase
space for the emission of extra particles. Kinematically, the
processes we consider fall into two broad categories, according to
whether the large scale which makes the process perturbative is the
virtuality of one of the incoming particles, 
as in deep-inelastic scattering (DIS), or the invariant mass of 
the final state, as in Drell-Yan (DY). 
We will not consider processes, such as jet or prompt-photon
production, in which the hard scale is provided by a transverse energy.
 As
is well known, the radiation of extra soft partons close to the
kinematic boundary enhances the the total cross section by
contributions  proportional to
powers of $\ln(1-x)$. These logs are generated by the integration over
the phase space of the (on-shell) emitted partons, and therefore, since
this phase space has an azimuthal symmetry, we expect an extra
emission in the final state to be accompanied by at most two powers of
the large log.

Consider specifically the Mellin moments of the cross section
\beq
\sigma(N,Q^2)\equiv \int_0^1 \!dx\, x^{N-1} \sigma(x,Q^2)
\label{meldef}
\eeq
and assume that for DIS-like processes they 
can be written in factorized form as
\bea
\sigma(N,Q^2)&=&C\left(N,Q^2/\mu^2,\as(\mu^2)\right)\,F(N,\mu^2)\nonumber\\
&=&C\left(N,1,\as(Q^2)\right)\,F(N,Q^2),
\label{pert}
\eea
where $F(N,\mu^2)$ are Mellin moments of a parton density
$F(x,\mu^2)$,
$C(N,Q^2/\mu^2,\as(\mu^2))$ is a hard perturbative coefficient, and
the last equation gives the standard renormalization-group improved
form of the factorized cross section.  In the case of Drell-Yan-like
processes, there are two partons in the initial state, and thus 
Eq.~(\ref{pert}) should be modified accordingly. 
In general, $F$ in Eq.~(\ref{pert}) is
an array of parton distributions (quarks and gluons), but the ensuing
complication of the formalism is trivial. Furthermore, to
next-to-leading order, as we will see explicitly in the sequel, there
is no quark-gluon mixing, so $F$ can be effectively taken to be a single parton
distribution.     
The validity of a factorization
of the form Eq.~(\ref{pert}) is the only assumption we will make on
the physical processes to which our resummation applies, and it
amounts to the assumption that the hard coefficient
$C\left(N,Q^2/\mu^2,\as(\mu^2)\right)$ can be multiplicatively
renormalized. We will discuss this in more detail in
Sect.~\ref{Resummation}.

Because resummation takes the form of an exponentiation, it is
convenient to classify large logs in terms of the log derivative of
the cross section $\sigma$, \ie, the so-called physical anomalous
dimension~\cite{physcat} defined as
\beq
\label{physad}
Q^2\frac{\partial\sigma(N,Q^2)}{\partial Q^2}=
\gamma(N,\as(Q^2))\,\sigma(N,Q^2).
\eeq
When addressing issues of operator mixing, it is convenient to choose a basis
of physical cross sections (or physical observables) equal in number
to the independent parton distributions, so the physical anomalous
dimension is a matrix.
The physical anomalous dimension $\gamma$ Eq.~(\ref{physad}) is
independent of factorization scale, and it is related
to the standard anomalous dimension $\gamma^{AP}$, defined by
\beq
\label{evolF}
\mu^2\frac{\partial F(N,\mu^2)}{\partial \mu^2}= 
\gamma^{\rm AP}(N,\as(\mu^2)) F(N,\mu^2),
\eeq
according to
\beq
\label{generic}
\gamma(N,\as(Q^2))
=\frac{\partial\ln C(N,Q^2/\mu^2,\as(\mu^2))}{\partial\ln Q^2}
=\gamma^{\rm AP}(N,\as(Q^2))
+\frac{\partial\ln C(N,1,\as(Q^2))}{\partial\ln Q^2}.
\eeq
It follows that the physical and standard anomalous
dimensions coincide at leading order, but differ beyond leading order.
In terms of the physical anomalous dimension, the cross section can be
written as 
\beq
\label{pert2}
\sigma(N,Q^2)=K(N;Q_0^2,Q^2)\,\sigma(N,Q_0^2)
=\exp\left[E(N;Q_0^2,Q^2)\right]\sigma(N,Q_0^2),
\eeq
where
\bea
\label{esplit} 
E(N;Q_0^2,Q^2)
&=&\int_{Q_0^2}^{Q^2}\frac{dk^2}{k^2}\gamma(N,\as(k^2))\label{edef}\\
&=&\int_{Q_0^2}^{Q^2}
\frac{dk^2}{k^2}\gamma^{\rm AP}(N,\as(k^2))
+\ln C(N,1,\as(Q^2))-\ln C(N,1,\as(Q_0^2)).
\nonumber
\eea

The structure of leading logs is then found by noting that upon Mellin
transformation  Eq.~(\ref{meldef}) the large $x$ region is mapped onto
the large $N$ region, so logs of $1-x$ become logs of $N$
and  conversely. Explicitly
\bea
\int_0^1 dx\,x^{N-1}\,\left[\frac{\ln^p(1-x)}{1-x}\right]_+
&=&\int_0^1 dx\,\frac{x^{N-1}-1}{1-x}\,\ln^p(1-x)\nonumber\\
&=&\frac{1}{p+1} \ln^{p+1} \frac{1}{N} -\gamma_E
\ln^{p}\frac{1}{N}+ O\left(\ln^{p-1} \frac{1}{N}\right),
\label{nlomel}
\eea
where $\gamma_E$ is the Euler gamma, \ie, $\gamma_E=-\psi(1)$. 
The resummed next-to-leading order
result of Refs.~\cite{cnt,sterman}  has the form
\beq
\label{ct2}
E^{\rm res}(N;Q_0^2,Q^2)=
a\int_0^1 dx\, \frac{x^{N-1}-1}{1-x}
\int_{Q_0^2(1-x)^a}^{Q^2(1-x)^a}
\frac{dk^2}{k^2}\hat g(\as(k^2))+ O(N^0)
\eeq
where
\beq
\label{gdef}
\hat g(\as)=\hat g_1\as+\hat g_2\as^2,
\eeq
and $a=1$ for processes with deep-inelastic kinematics, and $a=2$
for Drell-Yan kinematics.
This result implies that the physical anomalous dimension is given by
\bea
\gamma(N,\as(k^2))
&=&a\int_0^1 dx\, \frac{x^{N-1}-1}{1-x}\left[\hat g_1 \as(k^2(1-x)^a)
+\hat g_2 \as^2(k^2(1-x)^a)\right]
\nonumber\\
&&\quad\quad
+O\left(\as^{k+2}(k^2)\ln^k N\right)+ O(N^0)
\label{gamexp}
\\
&=&\gamma_1(\as(k^2) \ln N)+\as(k^2)\gamma_2(\as(k^2) \ln N)
+O\left(\as^{k+2}(k^2)\ln^k N\right)+ O(N^0),
\nonumber
\eea
where  $O(N^0)$ denotes non-logarithmic terms. 

This means that, up to terms which are finite as $N\to\infty$, the
physical anomalous dimension can be expanded as a power series in
$\as$ at fixed $\as\ln N$.  The leading log contribution to the
physical anomalous dimension is the sum of terms to all orders in
$\as$ where the powers of $\as$ and $\ln 1/N$ coincide; the
next-to-leading order terms have one more power of $\as$, and so on.
Because the QCD beta function starts at order $\as^2$, the order
$\as^k$ contribution to $\gamma$ Eq.~(\ref{gamexp}) leads to a
contribution of order $\as^k(Q^2)
\ln^{k+1} N$ to $E$ Eq.~(\ref{esplit}). 
In particular, the $O(\as)$ contribution corresponds to
a series of double-log contributions to the cross section (in $\ln
N$), in agreement with the simple power counting which leads to expect
at most two extra powers of $\ln N$ when the cross section is computed
at an extra order in $\as$.
Notice, however, that the power counting for the
physical anomalous dimension Eq.~(\ref{gamexp}) is in fact more
restrictive than what allowed on the basis of the fact that 
the cross section can contain at most double logs, in that it is free of
contributions of order $\as^{k+1}\ln^{2k} N$ with $k>1$. Indeed,
Eq.~(\ref{gamexp})  implies that the anomalous dimension contains at
most single logs, \ie, each extra power of $\as$ is accompanied at
most by an extra power of $\ln N$.

The resummed result Eq.~(\ref{gamexp}) 
implies that all the leading-log coefficients are
determined in terms of the single parameter $\hat g_1$, and the leading and
next-to-leading ones by $\hat g_1$ and $\hat g_2$: at leading log we have 
\beq
\gamma(N,\as(Q^2))
=a\hat g_1\int_0^1 dx\, \frac{x^{N-1}-1}{1-x}\,\as(Q^2)\sum_{j=0}^\infty 
(-\beta_0)^j\, \as^j(Q^2)\,\ln^j(1-x)^a,
\label{leadinglogct}
\eeq
where we have expanded the beta function as
\beq
\mu^2\frac{d}{d\mu^2}
\as(\mu^2)\equiv\beta(\as)
=-\beta_0\as^2(\mu^2)-\beta_1\as^3(\mu^2)+O(\as^4). 
\label{bet4}
\eeq
The term with $j=0$ in the series Eq.~(\ref{leadinglogct}) is a
contribution to the leading-order anomalous dimension: hence, $\hat
g_1$ is just the coefficient of the $\ln(1/N)$ contribution to the usual
one-loop anomalous dimension. Similarly, it is apparent that $\hat
g_2$ is the coefficient of the $\ln N$ contribution to the two-loop
anomalous dimension. 

If Eqs.~(\ref{ct2},\ref{gdef}) hold beyond
next-to-leading order, as claimed in Ref.~\cite{contopa}, \ie, if 
$\hat g(\as)$ is a power series in $\as$
with numerical coefficients $\hat g_k$, then $\hat g_k$ is just the
coefficient of the $\ln N$ contribution to the $k$-loop physical
anomalous dimension: the next$^k$-to-leading log
resummation is determined by knowledge of fixed-order $(k+1)$-loop
result.
In Sect.~\ref{RGI} we will prove a somewhat less
predictive form of the resummation, where the 
next$^k$-to-leading log resummation is determined  in general only 
by knowledge of a higher fixed-order computation.

Beyond leading order the standard anomalous dimension differs from the
physical one, so $\hat g_k$ with $k>1$ receives a contribution both from
the standard anomalous dimension and from the coefficient function.
It is thus natural to rewrite the resummation formula Eq.~(\ref{ct2})
in analogy to the unresummed separation Eq.~(\ref{esplit}) of
contributions to $E$ from the anomalous dimension and coefficient
function, by separating off the contribution which originates from the
anomalous dimension $\gamma^{\rm AP}$ Eq.~(\ref{evolF}):
\bea
&&E^{\rm res}(N;Q_0^2,Q^2)=
a\int_0^1 dx \frac{x^{N-1}-1}{1-x}
\Bigg[\int_{Q_0^2(1-x)^a}^{Q^2(1-x)^a}
\frac{dk^2}{k^2}A(\as(k^2))
\nonumber\\
&&\phantom{E^{\rm res}(N;Q_0^2,Q^2)=a\int_0^1 dx }
+B(\as(Q^2(1-x)^a))-B(\as(Q_0^2(1-x)^a))\,\Bigg]
\label{eressplit}
\eea
where $A(\as)=A_1\as +A_2\as^2$ 
is defined as the sum of one and two-loop
coefficients of $\ln(1/N)$ in the standard  anomalous dimension, and
\beq
B(\as)= B_1 \as
\label{bexp}
\eeq
makes up for the difference between physical and standard
anomalous dimensions:
\beq
\label{gAB}
\hat g(\as)=A(\as)+\frac{\partial B(\as(k^2))}{\partial\ln k^2}.
\eeq

We can then rewrite the resummed cross section 
\beq
\label{res2}
\sigma^{\rm res}(N,Q^2)=
\exp\left[E^{\rm res}(N;Q_0^2,Q^2)\right]\,\sigma^{\rm res}(N,Q_0^2)
\eeq
in factorized form according to Eq.~(\ref{pert})
by collecting all $Q^2$-dependent contributions to the resummation
Eq.~(\ref{eressplit}) into a resummed perturbative coefficient
$C^{\rm res}$ 
\bea
&&\sigma^{\rm res}(N,Q^2)=C^{\rm res}(N,Q^2/\mu^2,\as(\mu^2)) 
F(N,\mu^2)\label{ctres}\\
&&C^{\rm res}(N,Q^2/\mu^2,\as(\mu^2))\label{resfac}
\\
&&=
\exp\Bigg\{a\int_0^1 dx \frac{x^{N-1}-1}{1-x}
\Bigg[\int_{\mu^2}^{Q^2(1-x)^a}\frac{dk^2}{k^2}A(\as(k^2))
+B(\as(Q^2(1-x)^a))\Bigg]\Bigg\}.
\nonumber
\eea
The precise definition of the parton distribution $F$ and the
factorization scale $\mu^2$ will depend on the choice of factorization
scheme: according to the choice of scheme, the resummed terms will be
either part of the hard coefficient $C^{\rm res}$, or of the evolution
of the structure function $F$.
A detailed discussion of factorization scheme choices will be
given in Sect.~\ref{schemes}.

The resummed results Eq.~(\ref{ct2}), or
Eqs.~(\ref{ctres}-\ref{resfac}), can be explicitly cast in the form of
an exponentiation of logs of $1/N$. To this purpose, we note that
Eq.~(\ref{nlomel}) implies that, up to next-to-leading log level,
\bea
&&\int_0^1 dx\,\frac{x^{N-1}-1}{1-x}\,\ln^p(1-x)
=\frac{1}{p+1} \left(\ln\frac{1}{N}-\gamma_E\right)^{p+1}
\label{lnlnnl}\\
&&\phantom{\int_0^1 dx\,\frac{x^{N-1}-1}{1-x}\,\ln^p(1-x)}
=-\int_0^{1-\frac{1}{N}}\frac{dx}{1-x}     \left[1-\gamma_E
  \frac{d}{d\ln(1-x)} \right] \ln^p(1-x).
\nonumber 
\eea
The derivative term in the last expression is the first subleading
correction to the leading-log term. Note that, whereas a
next-to-leading log term can always be written as a derivative  of
the leading-log result, it is nontrivial that the coefficient of this
derivative in Eq.~(\ref{lnlnnl}) does  not depend on the power $p$ of
$\ln(1-x)$. This fact can be exploited to rewrite the resummed result
Eq.~(\ref{resfac}) as  
\bea
\ln C^{\rm res}(N,Q^2/\mu^2,\as(\mu^2))&=&-a\int_0^{1-\frac{1}{N}} 
\frac{dx}{1-x}
\left[\int_{\mu^2}^{Q^2(1-x)^a}
\frac{dk^2}{k^2}  A(\as(k^2))
+\tilde B(\as(Q^2(1-x)^a))\right]\nonumber\\
&=&-\int_{1}^{N^a} \frac{dn}{n} \left[\int_{n\mu^2}^{Q^2}
\frac{dk^2}{k^2} A(\as(k^2/n))
+\tilde B(\as(Q^2/n))\right],
\label{lnnlres}
\eea
where in the second step we have performed the change of variable
$n=(1-x)^{-a}$, and to this order $\tilde B(\as)=B(\as)- a\gamma_E
A_1\as$.  In Eq.~(\ref{lnnlres}) the exponentiation is directly
expressed in terms of logs of $1/N$.

Similarly, we can express the resummed physical anomalous dimension
Eq.~(\ref{ct2}) directly in terms of logs of $1/N$ by using
Eq.~(\ref{lnlnnl}) to perform the Mellin transform. We get
\bea
E^{\rm res}(N;Q_0^2,Q^2)&=&
a\int_0^{1-\frac{1}{N}} \frac{dx}{1-x}\, 
\int_{Q_0^2}^{Q^2}\frac{dk^2}{k^2}\,{g}(\as(k^2(1-x)^a))\label{ct4a}\\
&=&\int_1^{N^a} \frac{dn}{n}\, 
\int_{Q_0^2}^{Q^2}\frac{dk^2}{k^2}\,{g}\left(\as(k^2/n)\right),
\label{ct4}
\eea
where 
\beq
\label{ghatnl}
{g}(\as(\mu^2))=-\left(1-a\gamma_E\frac{d}{d\ln\mu^2}\right)\hat g(\as(\mu^2)).
\eeq

The physical anomalous dimension Eq.~(\ref{physad}) which 
corresponds to the resummed result
Eq.~(\ref{ct4}) is particularly simple: we get
\bea
\gamma(N,\as(Q^2))
&=& \int_1^{N^a} \frac{dn}{n}\,\left[g_1\as(Q^2/n)+g_2\as^2(Q^2/n)\right]
\label{newgamexp}\\
&=&
\frac{g_1}{\beta_0} \ln\frac{ \as(Q^2/N^a)}{\as(Q^2)}
+\frac{1}{\beta_0} \left(g_2-\frac{\beta_1}{\beta_0}
g_1\right) \left[\as(Q^2/N^a)-\as(Q^2)\right],
\label{newgamcalc}
\eea
where the coefficients of the two expansions Eq.~(\ref{gamexp}) and
Eq.~(\ref{newgamexp}) up to next-to-leading order are related by
Eq.~(\ref{ghatnl}):
\beq
g_1=-\hat g_1;\qquad  g_2=-\left( \hat g_2- a\gamma_E \beta_0 \hat g_1\right).
\label{coefrel}
\eeq
The leading- and next-to-leading-log coefficients in the expansion
of Eq.~(\ref{gamexp}) are given by
\bea
&&\gamma_1=-\frac{g_1}{\beta_0}
\ln\left[1+\beta_0\as(Q^2)\ln\frac{1}{N^a}\right]\label{gll}\\
&&\gamma_2=-\left(g_2-\frac{\beta_1}{\beta_0}g_1\right) 
\frac{\as(Q^2)\ln\frac{1}{N^a}}{1+\beta_0\as(Q^2)\ln\frac{1}{N^a}}
-\frac{g_1\beta_1}{\beta_0^2}
\frac{\ln\left[1+\beta_0\as(Q^2)\ln\frac{1}{N^a}\right]}
{1+\beta_0\as(Q^2)\ln\frac{1}{N^a}},
\label{gnll}
\eea
where we have used the beta function Eq.~(\ref{bet4}).

Summarizing, the resummation of Refs.~\cite{cnt,sterman} can be
compactly expressed in the two equivalent forms Eq.~(\ref{gamexp}) or
Eq.~(\ref{newgamexp}) of the resummed physical anomalous
dimension. The former shows that the resummed anomalous dimension is
the Mellin transform of a resummed leading-,
next-to-leading-,\dots,$\ln(1-x)$ splitting function, which in turn is
a power series in $\as(Q^2(1-x)^a)$ with numerical coefficients.  The
latter shows that this Mellin transform can be just expressed as a
series in $\as(Q^2/N^a)$, with numerical coefficients.  Knowledge of
the $k$-th coefficient of either expansion determines the
next$^{k-1}$-to-leading resummation. This coefficient can be read off
the coefficient of the $\ln N$ term in a fixed $k$-th order
computation of the physical anomalous dimension.
Up to next-to-leading order, the two forms of the resummation are related by
Eq.~(\ref{ghatnl}). 

Our aim here is to derive an all-order generalization of these
resummation formulae. A preliminary step is thus to obtain
an all-order generalization of the relation between the two available
forms of the next-to-leading log resummation.
This generalization will be derived in the next section.

\sect{$N$ space and $x$ space  beyond next-to-leading order}  
\label{Nx}

We wish to find an  all-order generalization of the relation
between leading $\ln N$ and leading $\ln(1-x)$ resummation discussed
in the previous section. The leading $\ln N$
resummation is expressed in terms
of
\beq
L_p\equiv\ln^p\frac{1}{N}=-p \int_0^{1-\frac{1}{N}} \frac{dx}{1-x} \ln^{p-1} (1-x),
\label{llndef}
\eeq
while the leading $\ln(1-x)$ resummation is expressed in terms of
\beq
I_p\equiv\int_0^1 dx\,\frac{x^{N-1}-1}{1-x}\,\ln^p(1-x).
\label{ipdef}
\eeq
In order to relate the two resummations we must therefore generalize
to all orders the next-to-leading order relation Eq.~(\ref{lnlnnl}),
\ie, compute the Mellin transform Eq.~(\ref{ipdef}) to all-order
logarithmic accuracy, up to power corrections.

To this purpose, we notice  that all integrals $I_p$ can be obtained from a
generating function $G(\eta)$:
\bea
I_p&=&\frac{d^p}{d\eta^p}\,G(\eta)\Bigg|_{\eta=0} ,\nonumber\\
G(\eta)&=&\int_0^1 dx\,(x^{N-1}-1)\,(1-x)^{\eta-1}.
\label{ipgen}
\eea
The integral can be determined at large $N$ using the Stirling
formula:
\beq
G(\eta)=\frac{\Gamma(N)\Gamma(\eta)}{\Gamma(N+\eta)}-\frac{1}{\eta}
=\frac{1}{\eta}\left[\frac{\Gamma(1+\eta)}{N^{\eta}}-1\right]
+ O\left(\frac{1}{N}\right).
\label{genfexp}
\eeq
On the other hand, the  generating function for $L_p$ is just $N^{-\eta}$:
\beq
L_p=\frac{d^p}{d\eta^p}N^{-\eta}\Bigg|_{\eta=0}.
\eeq
Hence, Eq.~(\ref{genfexp}) can 
be viewed as  a relation between the generating functions for
$I_p$ and $L_p$.
In particular, Taylor-expanding 
$\Gamma(1+\eta)$ in the expression Eq.~(\ref{genfexp}) of 
$G(\eta)$ 
leads to leading, next-to-leading, \dots $\ln N$ relations:
\bea
G(\eta)&=&\frac{1}{\eta}
\left
[\frac{1}{N^\eta}\sum_{k=0}^\infty \frac{\Gamma^{(k)}(1)}{k!}\eta^k-1\right]
\nonumber\\
&=&-\sum_{k=0}^\infty\frac{\Gamma^{(k)}(1)}{k!}\frac{d^k}{d\ln^k\frac{1}{N}}
\int_0^{1-\frac{1}{N}}\!dx\,(1-x)^{\eta-1}.
\label{gexp}
\eea

For the individual $I_p$ Eq.~(\ref{ipgen}) we get
\beq
\label{secondres}
I_p=-\sum_{k=0}^{p+1}
\frac{\Gamma^{(k)}(1)}{k!}
\frac{d^k}{d\ln^k\frac{1}{N}}\int_0^{1-\frac{1}{N}}dx\,\frac{\ln^p(1-x)}{1-x}
+ O\left(\frac{1}{N}\right).
\eeq
Because
\beq
\label{dll2}
\frac{d^k}{d\ln^k\frac{1}{N}}\int_0^{1-\frac{1}{N}}dx\,\frac{\ln^p(1-x)}{1-x}=
\int_0^{1-\frac{1}{N}}\frac{dx}{1-x}\,\frac{d^k\ln^p(1-x)}{d\ln^k(1-x)},
\eeq
Eq.~(\ref{secondres}) can be rewritten as
\beq
\label{secondres2}
I_p=-\sum_{k=0}^p
\frac{\Gamma^{(k)}(1)}{k!}
\int_0^{1-\frac{1}{N}}\frac{dx}{1-x}\,\frac{d^k}{d\ln^k(1-x)}\ln^p(1-x)
+ \Gamma^{(p+1)}(1)+ O\left(\frac{1}{N}\right).
\eeq
This is the desired generalization of Eq.~(\ref{lnlnnl}) to all orders.
Notice that the coefficient in the expansion remain $p$-independent
to all orders.
Performing the derivatives explicitly, we get
\beq
\label{firstres}
I_p=\frac{1}{p+1}\sum_{k=0}^{p+1}
\left(\begin{array}{c} p+1\\k \end{array}\right)\,\Gamma^{(k)}(1)\,
\left(\ln\frac{1}{N}\right)^{p+1-k}
+O\left(\frac{1}{N}\right).
\eeq

The inverse result, expressing $L_p$ in terms of $I_p$,
can be analogously found by inverting the relation between generating
functions: 
\beq
N^{-\eta}=\frac{\eta G(\eta)+1}{\Gamma(1+\eta)}.
\eeq
Proceeding as above, we get
\beq
\ln^n\frac{1}{N}=\sum_{k=1}^n
\frac{\Delta^{(k-1)}(1)}{(k-1)!}
\int_0^{1}dx\frac{x^{N-1}-1}{1-x}\,\frac{d^k}{d\ln^k(1-x)}\ln^n(1-x)
+\Delta^{(n)}(1)+ O\left(\frac{1}{N}\right),
\label{invsecondres2}
\eeq
where $\Delta^{(k)}(\eta)$ is the $k$-th derivative of
\beq
\Delta(\eta)\equiv\frac{1}{\Gamma(\eta)}.
\label{smallgdef}
\eeq
Evaluating the integrals explicitly we get
\beq
\label{invfirstres}
\ln^n\frac{1}{N}=\sum_{i=1}^n
\left(\begin{array}{c} n\\i \end{array}\right)\,\Delta^{(n-i)}(1)\,i\,I_{i-1}
+\Delta^{(n)}(1)+O\left(\frac{1}{N}\right).
\eeq

Because of the $p$-independence of  the coefficients of the expansion
Eq.~(\ref{secondres2}),  we can determine  explicitly the Mellin
transform of a generic
 function 
\beq
\hat g(\ln(1-x))=\sum_{p=0}^\infty \hat g_p \ln^p(1-x)
\label{genericfun}
\eeq
of $\ln(1-x)$, up to power corrections:
\bea
&&\int_0^1dx\,\frac{x^{N-1}-1}{1-x}\,\hat g(\ln(1-x))=
\sum_{p=0}^\infty \hat g_p\int_0^1dx\,\frac{x^{N-1}-1}{1-x}\,\ln^p(1-x)
\nonumber\\
&&\phantom{aaaaaaa}=
-\sum_{k=0}^\infty
\frac{\Gamma^{(k)}(1)}{k!}
\int_0^{1-\frac{1}{N}}\frac{dx}{1-x}\,\frac{d^k}{d\ln^k(1-x)}\hat g(\ln(1-x))
+O(N^0)\label{thirdres}\\
&&\phantom{aaaaaaa}=
\int_0^{1-\frac{1}{N}}\frac{dx}{1-x} g(\ln(1-x))
+O(N^0),\nonumber
\eea
where 
\beq
g (\ln(1-x))\equiv-\sum_{k=0}^\infty
\frac{\Gamma^{(k)}(1)}{k!}\,\frac{d^k}{d\ln^k(1-x)}\hat g(\ln(1-x)),
\label{dirfrel}
\eeq
and the last (constant) term in Eq.~(\ref{secondres2}) has been
dropped, so Eq.~(\ref{thirdres}) is  only correct up to non-logarithmic
terms, denoted by $O(N^0)$. 

The inverse relation can be analogously derived. Namely, we can cast
the integral of any function 
\beq
g(\ln(1-x))=\sum_{p=0}^\infty  g_p \ln^p(1-x)
\label{genericifun}
\eeq
as
a Mellin transform, up to non-logarithmic terms:
\bea
\int_0^{1-\frac{1}{N}}\frac{dx}{1-x} \, g(\ln(1-x))&=&
-\sum_{k=0}^\infty
\frac{\Delta^{(k)}(1)}{k!}
\int_0^1dx\,\frac{x^{N-1}-1}{1-x}\,\frac{d^k}{d\ln^k(1-x)}g(\ln(1-x))
+O(N^0)\nonumber\\
&=&\int_0^1dx\,\frac{x^{N-1}-1}{1-x}\,\hat g(\ln(1-x))
+O(N^0)\label{invthirdres}
\eea
where in the first step we have used Eq.~(\ref{invsecondres2}), and now
\beq
\hat g(\ln(1-x))\equiv-\sum_{k=0}^\infty
\frac{\Delta^{(k)}(1)}{k!}\,\frac{d^k}{d\ln^k(1-x)}g(\ln(1-x)).
\label{invfrel}
\eeq

Equations~(\ref{thirdres}-\ref{invfrel}) immediately provide us with
the sought-for all-order generalization of the relation between the
two forms of the resummed anomalous dimension Eq.~(\ref{gamexp}) and
Eq.~(\ref{newgamexp}), or of the resummation factor, Eqs.~(\ref{ct2})
and (\ref{ct4}). Specifically,
\beq
\label{ptogam}
\int_0^1 dx \frac{x^{N-1}-1}{1-x} \hat
g[\as(k^2(1-x)^a)]=\int_1^{N^a}\frac{dn}{n} g[\as(k^2/n)]
\eeq
with
\beq
\label{ghat}
g(\as(\mu^2))=-\sum_{p=0}^\infty \frac{\Gamma^{(p)}(1)a^p}{p!}
\frac{d^p}{d\ln^p\mu^2}\,\hat g(\as(\mu^2)).
\eeq
By choosing a factorization scheme, we can then obtain from the
resummation of the physical anomalous dimension a resummation of the
coefficient function \eg\ of the form of Eq.~(\ref{resfac}), with the
functions $A$ and $B$ computed to the corresponding order.

For future applications, it is interesting to observe that, through
similar manipulations, it is also possible to derive a relation
between a generic function of $\ln\frac{1}{N}$, and a function of
$\ln(1-x)$. Indeed, given a function
\beq
h\left(\ln\frac{1}{N}\right)=\sum_{p=0}^\infty h_p \ln^p\frac{1}{N}
\label{genericnfun}
\eeq
Eq.~(\ref{invsecondres2}) implies that
\beq
h\left(\ln\frac{1}{N}\right)=\sum_{k=1}^\infty
\frac{\Delta^{(k-1)}(1)}{(k-1)!}\int_0^1\!dx \frac{x^{N-1}-1}{1-x}
\frac{d^k}{d \ln^k(1-x)} h(\ln(1-x))
+O(N^0).
\label{llxtolln}
\eeq
The right-hand side of Eq.~(\ref{llxtolln}) can be viewed as the Mellin
transform of a function (more properly a distribution)
$\hat h(\ln(1-x))$:
\bea
&&h\left(\ln\frac{1}{N}\right)=\int_0^1\!dx\, x^{N-1}
\hat h(\ln(1-x));\nonumber\\
&&
\hat h(\ln(1-x))=
\sum_{k=1}^\infty
\frac{\Delta^{(k-1)}(1)}{(k-1)!}\left[\frac{1}{1-x}
\frac{d^k}{d \ln^k(1-x)} f\left(\ln(1-x)\right)\right]_++O(N^0).
\label{lxtolnmel}
\eea

\sect{All-order resummation}  
\label{RGI}

This section contains the main result of our paper, namely the proof
of an all-order generalization of the next-to-leading order
resummations discussed in Sect.~\ref{NLLres}. This proof exploits the
relation between $\ln (1-x)$ resummation and
$\ln N$ resummation established in Sect.~\ref{Nx}, and  
proceeds in two steps. First, (in
Sect.~\ref{proof}) we determine the $N$ dependence of the regularized
coefficient function in the large-$N$ limit. We show that 
the kinematic
structure of the $k$-particle phase
space essentially determines the form of the
$N$-dependence in the large $N$ limit, and that loop integrations
do not modify this result.
Then (in
Sect.~\ref{Resummation}) we prove that, given this
form of the $N$-dependence of the
regularized  cross section, renormalization group
invariance fixes the all--order dependence of the 
physical anomalous dimension Eq.~(\ref{physad}) in such a way that
an infinite class of
leading, next-to-leading,\dots resummations are found in terms of
corresponding fixed order results. In particular, in
Sect.~\ref{nlores}
we show that the
next-to-leading-log resummation Eq.~(\ref{ct4}) can be  obtained  by
exploiting available fixed-order results.

\subsection{Kinematic structure of soft logs}
\label{proof}

We consider the perturbative expansion of the bare coefficient
function in powers of the bare coupling constant $\az$,
\beq
C^{(0)}(x,Q^2,\az,\epsilon)=\sum_{n=0}^\infty \az^n\,
C^{(0)}_n(x,Q^2,\epsilon)
\eeq
and its Mellin transform
\beq
C^{(0)}(N,Q^2,\az,\epsilon)
=\sum_{n=0}^\infty \az^n\,C^{(0)}_n(N,Q^2,\epsilon)
\label{Cbare}
\eeq
in dimensional regularization with $d=4-2\epsilon$ space-time
dimensions. By coefficient function here we mean a vector
whose elements are the parton-level cross section for the partonic
subprocesses that contribute to the given process: the formalism
and the results of this section remain true even in the presence of
parton mixing.
 
The result that will be proven in this section, and then
renormalization-group improved in the next section, is that in the large-$N$
limit, $C^{(0)}_n\left(N,Q^2,\epsilon\right)$ has the following structure:
\beq
C^{(0)}_n(N,Q^2,\epsilon)
=\sum_{k=0}^n C^{(0)}_{nk}(\epsilon)\,
(Q^2)^{-(n-k)\epsilon}\,
\left(\frac{Q^2}{N^a}\right)^{-k\epsilon}+O(1/N),
\label{salame}
\eeq
where $O(1/N)$ denotes all terms which vanish as $N\to\infty$ in the
limit $\epsilon\to0$. We will also see that the coefficients
$C^{(0)}_{nk}$ have a pole of order $2n$ in $\epsilon$, related to
infrared singularities. Even though such poles cancel in the
coefficient function, which is free of infrared
singularities~\cite{ircan}, the interference of the poles with the powers of
$N^{-\epsilon}$ is responsible for the presence of powers of $\ln N$
in the renormalized four-dimensional cross section. Therefore, soft
logs may be viewed as being due to an incomplete cancellation between
real and virtual contributions to $C^{(0)}_n$ in the soft limit.

Equation~(\ref{salame}) will be established by proving that
\beq
C^{(0)}_n(x,Q^2,\epsilon)=(Q^2)^{-n\epsilon}\,
\left[C^{(0)}_{n0}(\epsilon)\,\delta(1-x)
+\sum_{k=1}^n \frac{C^{(0)}_{nk}(\epsilon)}{\Gamma(-ak\epsilon)}\,
(1-x)^{-1-ak\epsilon}\right]+O\left[(1-x)^0\right] ,
\label{salamex}
\eeq
where $O\left[(1-x)^0\right]$ denote terms which are not divergent
as $x\to1$ in the $\epsilon\to0$ limit. 
Eq.~(\ref{salamex}) implies Eq.~(\ref{salame}) because
\beq
\int_0^1 dx\,x^{N-1}\,(1-x)^{-1-ak\epsilon}=
\frac{\Gamma(N)\Gamma(-ak\epsilon)}{\Gamma(N-ak\epsilon)}
=\Gamma(-ak\epsilon)\,N^{ak\epsilon}+ O(1/N),
\label{mellinst}
\eeq
while the Mellin transform of any function of $x$ which is not divergent
as $x\to1$ (\eg\ any non-negative power of $1-x$)
vanishes as $N\to\infty$.

The content of Eq.~(\ref{salame}) is that, in the large $N$ limit,  
the dependence of the
regularized cross section on $N$ only goes
through integer powers of the dimensionful variable
$\left(Q^2/N^a\right)^{-\epsilon}$. Because in $d$ dimensions the
coupling constant has the dimensions of $Q^{2\epsilon}$, the
dependence on this dimensionful variable is related through the
renormalization group to the running of the coupling. As we shall show
in Sect.~\ref{Resummation}, this is sufficient to establish the
desired all-order resummation. In the remainder of this subsection we
will show that Eq.~(\ref{salamex}) essentially follows from the
kinematics of soft  emission. The reader who is not interested
in technicalities  can skip directly to
Sect.~\ref{Resummation} where only the result Eq.~(\ref{salame}) will be used.
 
We consider deep-inelastic scattering and Drell-Yan production as 
representative cases of the two classes of processes discussed in
Sect.~\ref{NLLres}. 
In the case of deep-inelastic scattering, the
relevant parton subprocesses are
\bea
\gamma^*(q)+{\cal Q}(p)&\to& {\cal Q}(p')+{\cal X}
\label{qgamma}
\\
\gamma^*(q)+{\cal G}(p)&\to& {\cal Q}(p')+{\cal X},
\label{ggamma}
\eea
where $\cal Q$ is a quark or an antiquark, $\cal G$ a gluon, and $\cal
X$ is any collection of quarks and gluons.   In the case of Drell-Yan, 
the relevant subprocesses are
\bea
{\cal Q}(p)+{\cal Q}(p')&\to& \gamma^*(q)+{\cal X}
\label{qq}
\\
\label{qg}
{\cal Q}(p)+{\cal G}(p')&\to& \gamma^*(q)+{\cal X}
\\\label{gg}
{\cal G}(p)+{\cal G}(p')&\to& \gamma^*(q)+{\cal X}.
\eea

We are interested in the most singular
contributions to the cross section as $x\to1$, and specifically in the
way these terms depend on $1-x$. To this purpose, we consider  the
dependence on $1-x$ of the phase space and of the amplitude for the
generic processes Eq.~(\ref{qgamma}-\ref{gg}). The structure of the
amplitude will in turn be discussed by considering first tree-level
processes, and then processes with loops. We will first carry out the
proof for deep-inelastic scattering, and then for Drell-Yan.

As discussed in the Appendix, the phase space for a generic process
with incoming momentum $P$ and $n$ bodies  in the final state 
with outgoing momenta $k_1$,\dots,$k_n$ can be expressed in terms
 of two-body phase space integrals by using recursively the identity
Eq.~(\ref{ntonmo}).
For  DIS-like processes, all outgoing particles are massless partons.
Indicating with $p$ and $p'$ the incoming and outgoing quark
momenta and with $q$ the $\gamma^*$ momentum we have 
\beq
d\phi_{n+1}(p+q;k_1,\ldots,k_n,p')=
\int_0^{s}\frac{dM_n^2}{2\pi}\,
d\phi_2(p+q;k_n,P_n)\,d\phi_{n}(P_n;k_1,\ldots,k_{n-1},p'),
\label{ggdis}
\eeq
where
\beq
s=(p+q)^2=\frac{Q^2(1-x)}{x}.
\label{sdis}
\eeq
Applying Eq.~(\ref{ggdis}) recursively we get 
\bea
d\phi_{n+1}(P_{n+1};k_1,\ldots,k_n,p')&=&
\int_0^{s}\frac{dM_n^2}{2\pi}\,
d\phi_2(p+q;k_n,P_n)\,
\int_0^{M_n^2}\frac{dM_{n-1}^2}{2\pi}\,
d\phi_2(P_n;k_{n-1},P_{n-1})\,
\nonumber\\
&&\ldots
\,\int_0^{M_3^2}\frac{dM_2^2}{2\pi}\,
d\phi_2(P_3;k_2,P_2)\,
d\phi_2(P_2;k_1,P_1),
\label{ggr}
\eea
where we have defined $P_{n+1}\equiv p+q$, so  $M^2_{n+1}=s$, and
$P_1\equiv p'$, so  $P_1^2=0$.

The dependence of the phase space on $1-x$ can now be traced by
performing the change of variables
\beq
z_i=\frac{M_i^2}{M_{i+1}^2};\quad M_i^2=sz_n\ldots z_i;\quad i=2,\ldots n.
\label{zdefdis}
\eeq
All $z_i$ range between $0$ and $1$, and
\beq
dM_n^2\ldots dM_2^2 = s^{n-1}z_n^{n-2}z_{n-1}^{n-3}\ldots z_3\,
dz_n\ldots dz_2.
\eeq
The two-body phase space Eq.~(\ref{ftbs})  is then
\beq
d\phi_2(P_{i+1};k_i,P_i)=N(\epsilon)\,
s^{-\epsilon}\,
\left(z_n\, z_{n-1}\ldots z_{i+1}\right)^{-\epsilon}\,
(1-z_i)^{1-2\epsilon}\,d\Omega_i,
\label{twobodyc}
\eeq
where $d\Omega_i$ is the angular integral in the center-of-mass frame
of the $(k_i,\, P_i)$ system, and
$N(\epsilon)=\frac{1}{2(4\pi)^{2-2\epsilon}}$.

Hence, the $n$-body phase space  Eq.~(\ref{ggr}) can be rewritten as
\bea
&&d\phi_{n+1}(p+q;k_1,\ldots,k_n,p')=
2\pi\left[\frac{N(\epsilon)}{2\pi}\right]^n
s^{n-1-n\epsilon}\,d\Omega_n\ldots d\Omega_1
\nonumber\\
&&\qquad\qquad\int_0^1 dz_n\,z_n^{(n-2)-(n-1)\epsilon}\,(1-z_n)^{1-2\epsilon}
\ldots
\int_0^1 dz_2\,z_2^{-\epsilon}\,(1-z_2)^{1-2\epsilon}.
\label{ggr2}
\eea
The dependence of the phase space on $1-x$ comes
entirely from the prefactor of $s^{n-1-n\epsilon}$: the dependence on
$x$ and $Q^2$ has been entirely removed from the integration
range. 
Now, the amplitude whose square modulus  is integrated with the phase space
Eq.~(\ref{ggr2}) is in general a function
$A_{n+1}(Q^2,s;z_2,\dots,z_n;\Omega_1,\dots,\Omega_n)$: it 
depends on $3(n+3)-10+1=3n$ independent variables
for a process with two initial-state
and $n+1$ final-state particles, in which one of the masses (the
virtuality $Q^2$) is not fixed. In the $x\to1$ limit, $s\to0$ and
the dominant contribution is given by terms which are most singular as
$s$ vanishes. Because of the cancellation of infrared
singularities~\cite{ircan}, $|A_{n+1}|^2\tozero{s}
s^{-n+O(\epsilon)}$: 
a stronger singularity would lead to powerlike infrared
divergences. Less singular terms instead correspond to contributions
whose Mellin transform vanishes in the large-$N$ limit. 

Hence only
terms in the square amplitude which behave as $s^{-n+O(\epsilon)}$ 
contribute in
the $x\to1$ limit. In $d$ dimensions, these terms pick up an 
$s^{-n\epsilon}$ prefactor from the phase space Eq.~(\ref{ggr2}). In
order to establish the result Eq.~(\ref{salamex}) we must study the
dependence of these terms on $s^\epsilon$. The case of
a tree-level amplitude, \ie, purely real soft emission is in
fact straightforward. Indeed,  
in this case, the amplitude manifestly does not contain any power of
$s^\epsilon$, so the result  Eq.~(\ref{salamex}) follows immediately,
with $a=1$. 
Notice that in this case, since the factors of $s^{-\epsilon}$ come from
the phase space only, 
the sum over $k$ in Eq.~(\ref{salamex}) reduces to the single term
$k=n$: the power of $\left[Q^2(1-x)\right]^{-\epsilon}$ coincides with 
the number of $z$ integrations, \ie, with the number of emitted partons. 

To understand the physical meaning of this result, note that in the 
center-of-mass frame
of the $k_i,\, P_i$ system the energy of the $i$-th of the $n$ emitted
partons  is (see Appendix~\ref{phsp})
\beq
\omega_i=k^0_i=\frac{M_{i+1}}{2}
\left(1-\frac{M_i^2}{M_{i+1}^2}\right).
\label{3momi}
\eeq
Now, concentrate on the emission of the  $n$-th parton, whose
energy is
$k^0_n=\frac{\sqrt{s}}{2}(1-z_n)=
\frac{\sqrt{s}}{2}\left(1-\frac{M^2_{i}}{s}\right)$.
The singularity arises 
if the squared amplitude behaves as $1/(k^0_n)^2$. This behaviour is seen
to be present \eg\ for gluon emission in the eikonal
approximation. In such case we get
\bea
&& |A_{n+1}|^2d\phi_{n+1}\sim
\int_0^{k^0_{\rm
max}}\frac{d{k^0_n}}{{k^0_n}^{1+2\epsilon}}\,d\phi_{n}\label{logint}
=\frac{s^{-\epsilon}}{4}\int_0^1\frac{dz_n}{(1-z_n)^{1+
2\epsilon}}\,d\phi_{n}
\\&&\qquad\qquad=
-\frac{1}{2\epsilon}\left[\frac{Q^2(1-x)}{4x}\right]^{-\epsilon}\,d\phi_{n}
\label{sloppy}
\eea
where the maximum value of ${k^0_n}$ is attained when $M^2_n=0$ and is
\beq
k^0_{\rm
max}=\frac{\sqrt{s}}{2}=\sqrt{\frac{Q^2(1-x)}{4x}}.
\label{kmax}
\eeq
The
$x$ dependence of the result is thus simply determined kinematically
in terms of  the upper limit of the logarithmic ${k^0_n}$ integration.
Eq.~(\ref{ggr2}) then shows that, thanks to the structure of the
$n$-body phase space, this remains true in the general case of
$n$-parton emission. 

We will now study how this result is
modified by the inclusion of loops. To this purpose, we notice that 
a generic amplitude with loops can be viewed as a
tree-level amplitude formed with proper vertices. Powers of
$s^\epsilon$ can only arise from loop integrations in the proper
vertices.  We thus consider
only purely scalar loop integrals, since numerators of fermion
or vector propagators and vertex factors cannot induce any dependence on
$s^{\epsilon}$. 

 Let us therefore consider an
arbitrary proper diagram $G$ in a massless scalar theory with $E$
external lines, $I$
internal lines and $V$ vertices.
It can be shown~\cite{IZ} that, denoting with $P$ the set of 
$E$ external momenta and $P_E$ the set of  independent invariants 
  the corresponding amplitude
$\tilde{A}_G(P_E)$ has the form
\bea 
&&\tilde{A}_G(P_E)=K\,(2\pi)^d\,\delta^{(d)}(\sum P) A_G(P_E)\\
\nonumber
&&\qquad A_G(P_E)=\frac{i^{I-L(d-1)}}{(4\pi)^{dL/2}}
\Gamma(I-dL/2)\prod_{l=1}^I
\left[\int_0^1 d\beta_l\right]
\frac{\delta(1-\sum_{l=1}^I \beta_l)}{[P_G(\beta)]^{d(L+1)/2-I}
\left[D_G(\beta,P_E)\right]^{I-dL/2}},
\label{G5}
\eea
In Eq.~(\ref{G5}),  $\beta_l$ are the usual
Feynman parameters, $P_G(\beta)$ is a homogeneous
polynomial of degree $L$ in the $\beta_l$,  $D_G(\beta,P_E)$ is a
homogeneous polynomial of degree $L+1$ in the $\beta_l$ with
coefficients which are linear functions of the scalar products
of the set $P_E$, \ie\ with dimensions of (mass)$^2$, 
and $K$ collects all overall factors, such as
couplings and symmetry factors.

The amplitude  $\tilde{A}_G(P_E)$ Eq.~(\ref{G5}) depends on $s$ only through 
$D_G(\beta,P_E)$, which, in turn, is linear in $s$. We can then
determine in general the dependence of $\tilde{A}_G(P_E)$ 
by considering two possible cases. The first possibility
is that $D_G(\beta,P_E)$ is independent of
all invariants except $s$, \ie\ $D_G(\beta,P_E)=s d_G(\beta)$. In such case, 
$A_G(P_E)$ depends on $s$ as
\beq
 A_G(P_E)=\left(\frac{1}{s}\right)^{I-2L+L\epsilon}  a_G,
\label{G6}
\eeq
where $a_G$ is a numerical constant, obtained performing the Feynman
parameter integrals. The second possibility is that $D_G(\beta,P_E)$
depends on some of the other invariants.
In such case, $A_G(P_E)$ is manifestly an analytic function of $s$ at
$s=0$, and thus it can be expanded in  Taylor series around $s=0$,
with coefficients which depend on the other invariants.
In the former case,
Eq.~(\ref{G6}) implies that the $s$ dependence induced by
loop integrations in the square amplitude  is given by 
integer powers of $s^{-\epsilon}$. In the latter case, the $s$ 
dependence induced by
loop integrations in the square amplitude is given by integer positive
powers of $s$. 

Therefore, we conclude that 
Eq.~(\ref{salamex}) holds in general even when loop integrations are present.
Unlike in the case of tree-level diagrams, however,
the overall power of  $s^{-\epsilon}$ is no longer determined by
the phase space only, and thus no longer equal to the
number of emitted partons, from which it can differ by an integer
amount. Specifically, each loop integration can carry at most a factor
of $s^{-\epsilon}$. Hence, at order $\alpha_0^n$ the highest power of
$s^{-\epsilon}$ is $n$, given that either an extra real emission or an
extra loop can give at most an extra factor $s^{-\epsilon}$, but
otherwise contributions proportional to all integer powers of
$s^{-\epsilon}$
up to $n$ are allowed. 
 Because in $d$ dimensions the bare coupling constant
$\alpha_0$ has mass dimensions $2\epsilon$, the dependence
on $Q^2$, which is the only other dimensionful variable, is fixed by
dimensional analysis, and Eq.~(\ref{salamex}) immediately follows.

Furthermore, it is clear (compare  Eq.~(\ref{sloppy})) that each
logarithmic $z_i$ integration produces a $1/\epsilon$ pole. Since
each angular integration gives a $1/\epsilon$ pole from the collinear
region, the highest soft singularity due to real emission in the
coefficient $C_{nk}^{(0)}/\Gamma(-a k \epsilon)$ in Eq.~(\ref{salamex}) is an
$\epsilon$ pole of order $2k-1$.
Hence, $C_n^{(0)}$ has a soft singularity 
$1/\epsilon^{2n}$, from the purely real emission term 
$C^{(0)}_{nn}$  in the sum
Eq.~(\ref{salame}).
Because infrared singularities must cancel~\cite{ircan} in
$C_n^{(0)}$, the sum of all
other terms $C^{(0)}_{nk}$ in the sum also has an order $2n$ pole
in $\epsilon$.

The whole argument can be reproduced for Drell-Yan-like processes with
minor modifications which account for the different
kinematics. Specifically, in the case of Drell-Yan 
we have massless partons in the initial state 
with momenta $p,\,p'$, while the  Drell-Yan pair in the final state
has momentum $Q$, so  $x=Q^2/s$,  with $s=(p+p')^2$. The recursive
expression for the $n$-body phase space Eq.~(\ref{ntonmo})
now becomes
\beq
d\phi_{n+1}(p+p';Q,k_1,\ldots,k_n)=
\int_{Q^2}^{s}\frac{dM_n^2}{2\pi}\,
d\phi_2(p+p';k_n,P_n)\,d\phi_{n}(P_n;Q,k_1,\ldots,k_{n-1}),
\label{ggdy}
\eeq
leading to
\bea
d\phi_{n+1}(P_{n+1};Q,k_1,\ldots,k_n)&=&
\int_{Q^2}^{s}\frac{dM_n^2}{2\pi}\,d\phi_2(p+p';k_n,P_n)\,
\int_{Q^2}^{M_n^2}\frac{dM_{n-1}^2}{2\pi}\,
d\phi_2(P_n;k_{n-1},P_{n-1})\,
\nonumber\\
&&\ldots
\,\int_{Q^2}^{M_3^2}\frac{dM_2^2}{2\pi}\,
d\phi_2(P_3;k_2,P_2)\,
d\phi_2(P_2;k_1,P_1),
\label{ggrdy}
\eea
where now  $P_{n+1}\equiv p+p'$, so  $M^2_{n+1}=s$, and
$P_1\equiv Q$.

The change of variables which separates off the dependence on $1-x$ is
now
\beq
z_i=\frac{M_i^2-Q^2}{M_{i+1}^2-Q^2}; \quad M_i^2-Q^2=(s-Q^2)z_n\ldots
z_i;\quad i=2,\ldots n 
\label{zdefdy}
\eeq
so all $z_i$ still range between $0$ and $1$ and
\beq
dM_n^2\ldots dM_2^2 = (s-Q^2)^{n-1}z_n^{n-2}z_{n-1}^{n-3}\ldots z_3\,
dz_n\ldots dz_2.
\eeq
The two-body phase space Eq.~(\ref{ftbs}) is now
\bea
d\phi_2(P_{i+1};k_i,P_i)&=&N(\epsilon)\,(M_{i+1}^2)^{-1+\epsilon}\,
\left[(M_{i+1}^2-Q^2)-(M_i^2-Q^2)\right]^{1-2\epsilon}\,d\Omega_i
\nonumber\\
&=&N(\epsilon)\,(Q^2)^{-1+\epsilon}\,
(s-Q^2)^{1-2\epsilon}(z_n\ldots z_{i+1})^{1-2\epsilon}
(1-z_i)^{1-2\epsilon}\,d\Omega_i,
\eea
where in the last step we have replaced  $(M_{i+1}^2)^{-1+\epsilon}$
by $(Q^2)^{-1+\epsilon}$ in the  $x\to 1$ limit
(compare Eq.~(\ref{zdefdy})). 

We finally get
\bea
&&d\phi_{n+1}(p+q;k_1,\ldots,k_n,p')\nonumber\\
&&\qquad=
2\pi\left[\frac{N(\epsilon)}{2\pi}\right]^n\,(Q^2)^{-n(1-\epsilon)}
(s-Q^2)^{2n-1-2n\epsilon}\,
d\Omega_n\ldots d\Omega_1
\nonumber\\
&&\qquad\int_0^1 dz_n\,z_n^{(n-2)+(n-1)(1-2\epsilon)}\,(1-z_n)^{1-2\epsilon}
\ldots
\int_0^1 dz_2\,z_2^{1-2\epsilon}(1-z_2)^{1-2\epsilon}.
\label{ggr2dy}
\eea
The dependence on $1-x$ is now entirely contained in the prefactor
\beq
(Q^2)^{-n(1-\epsilon)}(s-Q^2)^{2n-1-2n\epsilon)}
=
\frac{x^{1-2n+2n\epsilon}}{Q^2(1-x)}\,[Q^2(1-x)^2]^{n-n\epsilon}.
\eeq

Again, this proves the result Eq.~(\ref{salamex}) 
for real (tree-level) emission,
since the corresponding amplitude is free of factors of
$(s-Q^2)^{-\epsilon}$. Note however that now $a=2$. 
The physical interpretation is the same as in
the deep-inelastic case, except that the upper limit of the
logarithmic $k_n^0$ integration Eq.~(\ref{logint}) for the $i$-th parton
is now attained when $M^2_i=Q^2$ and is
\beq
k^0_{\rm
max}=\frac{\sqrt{s}}{2}\left(1-\frac{Q^2}{s}\right)=
\sqrt{\frac{Q^2(1-x)^2}{4x}}. 
\label{kmaxdy}
\eeq
Finally, loops can be included as in the case of deep-inelastic
processes, with the  possible cut at $x=1$ now being given by a factor of 
$(s-Q^2)^{I-2L+L\epsilon}$.

We conclude that Eq.~(\ref{salamex}) holds for both deep-inelastic and
Drell-Yan-like processes, with  $a=1$ in the former case and  $a=2$
in the latter. This difference is
merely a reflection of the underlying
kinematics, Eqs.~(\ref{kmax}) and~(\ref{kmaxdy}), but in both cases
the soft logs (\ie\ powers of $(1-x)^{-\epsilon}$) can be traced to the
fact that the phase space variables $M^2_i$ can all
be related to a single dimensionful variable $s$ (DIS) or $s-Q^2$ (DY)
through dimensionless variables $z_i$ which range from $0$ to $1$,
according to Eqs.~(\ref{zdefdis},\ref{zdefdy}). 

It is interesting to ask which are  the
kinematic configurations which contribute to the $x\to1$ limit: in
fact, they turn out not no be the  same in deep-inelastic and Drell-Yan. 
Indeed, consider a contribution to the cross
section that involves $n$ massless partons with momenta
$k_1,\ldots,k_n$ in the final state. We have
\beq
\label{DYmom}
p+p'=q+k_1+ \ldots+k_n.
\eeq
Squaring both sides of Eq.~(\ref{DYmom}) we get
\beq
s(1-x)=\sum_{i,j=1}^n k_i\cdot  k_j+2\sum_{i=1}^n q\cdot k_i.
\eeq
The quantities
\beq
q\cdot k_i = k^0_i\left(q_0-\sqrt{q_0^2-Q^2}\cos\theta_i\right)
\eeq
are positive semi-definite, and vanish only for $k_i^0=0$. Hence, for
Drell-Yan processes, when $x$ approaches 1, all $k^0_i$'s must go to
zero. \ie\ all emitted partons must be soft.

On the other hand, in the case of DIS we have
\beq
p+q=k_1+\ldots+k_n
\eeq
(where $k_1=p'$), which gives
\beq
\frac{Q^2(1-x)}{x}=\sum_{i,j=1}^n k_i\cdot  k_j
=\sum_{i,j=1}^n k^0_i  k^0_j\, (1-\cos\theta_{ij}),
\label{momconsdis}
\eeq
where $\theta_{ij}$ is the angle between $\vec{k}_i$ and $\vec{k}_j$.
Now take the limit $x\to 1$. The most general solution to
Eq.~(\ref{momconsdis}) is not $k^0_i=0$ for all $i$, but rather
\beq
\label{sol}
\begin{array}{ll}
k_i^0=0 \quad & {\rm for}\quad 1\leq i\leq \bar{n}
\\
\theta_{ij}=0; k^0_i,k^0_j\;
{\rm arbitrary} & {\rm for} \quad \bar{n}\le i,j \leq n.
\end{array}
\eeq
In other words, for DIS the kinematical configurations that correspond to
$x=1$ include not only the case when all partons in the final state
have small energies, but also the cases when a subset of them are
collinear to each other.

\subsection{Renormalization group improvement}
\label{Resummation}
We now study the restrictions that renormalization--group invariance
imposes on the cross section. As already mentioned, our only
assumption is the standard factorization Eq.~(\ref{pert}), namely,
that $C(N, Q^2/\mu^2,\as(\mu^2))$ can be multiplicatively
renormalized.  This means that all divergences can be removed from the
bare coefficient $C^{(0)}\left(N,Q^2,\az,\epsilon\right)$
Eq.~(\ref{Cbare}) by defining a renormalized coupling $\as(\mu^2)$
according to the implicit equation
\beq
\az(\mu^2,\as(\mu^2),\epsilon)=
\mu^{2\epsilon}\, \as(\mu^2)\,\zalpha
\label{alphabare}
\eeq
and a renormalized coefficient
\beq
C\left(N,Q^2/\mu^2,\as(\mu^2),\epsilon\right)
=\zc\, C^{(0)}\left(N,Q^2,\az,\epsilon\right),
\label{mulnor}
\eeq
where $\mu$ is the renormalization scale and $\zalpha$ and $\zc$ are
computable in perturbation theory and have multiple poles at
$\epsilon=0$. The renormalized coefficient function
$C\left(N,Q^2/\mu^2,\as(\mu^2),\epsilon\right)$ is finite at
$\epsilon=0$, and, because the renormalized coupling $\alpha_s$ is
dimensionless, it can only depend on $Q^2$ through $Q^2/\mu^2$.

In a mass-independent renormalization (and factorization) scheme such
as \MS, both $Z^{(\as)}$ and $Z^{(C)}$ are series in $\as$ with
$Q^2$--independent coefficients.  Therefore, 
the physical anomalous dimension is
\beq
\gamma\equiv Q^2\frac{d}{dQ^2}\ln C\left(N,Q^2/\mu^2,\as,\epsilon\right) =
-\epsilon \az\frac{\partial}{\partial\az} \ln 
C^{(0)}\left(N,Q^2,\az,\epsilon\right),
\label{gcalc}
\eeq
where we have exploited the fact that, for dimensional reasons,
$C^{(0)}$ can only depend on $Q^2$ through the combination
$\az Q^{-2\epsilon}$. \footnote{Note that the standard anomalous dimension
$\gamma^{\rm AP}$ Eq.~(\ref{evolF}) is instead given by
$$\gamma^{\rm AP}(N,\as(\mu^2))=-
\frac{\partial\ln Z^{(C)}(N,\as(\mu^2),\epsilon)}
{\partial\ln\mu^2},$$
due to the fact that the bare coefficient $C^{(0)}$ is independent of
$\mu^2$. It follows that $\gamma^{\rm AP}=\gamma$ if and only if $\zalpha=1$,
\ie\ if the four--dimensional beta function vanishes.}
Equation (\ref{gcalc}) implies that in $d$ dimensions
the physical anomalous dimension $\gamma$
viewed as a function of $\az$ admits an expansion in powers of
$\az Q^{-2\epsilon}$, while as a function of $\as(\mu^2)$ it admits an
expansion in powers of $(Q^2/\mu^2)^{-\epsilon}$. 

In the large $N$ limit, the dependence of the bare coefficient function on
$N$ is given by Eq.~(\ref{salame}). But Eq.~(\ref{gcalc}) 
implies that $\gamma$ has the same property, namely
\bea
\gamma\left(N,Q^2,\az,\epsilon\right)
&=&\sum_{i=1}^{\infty}\az^i 
\sum_{j=0}^i\gamma_{ij}(\epsilon)\,(Q^2)^{-(i-j)\epsilon}
\left(\frac{Q^2}{N^a}\right)^{-j\epsilon}+O\left(\frac{1}{N}\right)
\nonumber\\
&=&\sum_{i=1}^{\infty}\sum_{j=0}^i \gamma_{ij}(\epsilon) \,
\left[(Q^2)^{-\epsilon}\az\right]^{i-j} 
\left[\left(\frac{Q^2}{N^a}\right)^{-\epsilon}\az\right]^j
+O\left(\frac{1}{N}\right).
\label{ldef}
\eea
Hence, in the large  $N$ limit the $d$-dimensional physical 
anomalous dimension is a power series in
$Q^{-2\epsilon}\az$ and 
$(Q^2/N^a)^{-\epsilon}\az$,
with $N$--independent coefficients.

The renormalized expression of the physical anomalous dimension is
found expressing in Eq.~(\ref{ldef}) 
the bare coupling in terms of the renormalized one by means
of  Eq.~(\ref{alphabare}): this yields 
an expression of the physical anomalous dimension which is finite in the limit
$\epsilon\to0$.
Now, the function 
\beq
\bar\alpha_0(Q^2/\mu^2,\as(\mu^2),\epsilon)=
Q^{-2\epsilon}\az(\mu^2,\as(\mu^2),\epsilon)
=\left(\frac{Q^2}{\mu^2}\right)^{-\epsilon}\, \as(\mu^2)\,\zalpha
\label{alphabar}
\eeq
is manifestly renormalization-group invariant:
\beq
\mu^2 \frac{d\bar\alpha_0}{d\mu^2} =0.
\label{abrgi}
\eeq
It follows that
\beq
\bar\alpha_0(Q^2/\mu^2,\as(\mu^2),\epsilon)=
\bar\alpha_0(1,\as(Q^2),\epsilon)
=\as(Q^2)\,\zalphaQ.
\label{renal}
\eeq
Explicitly, noting that the $d$ dimensional beta function is given by
\bea
\mu^2\frac{d}{d\mu^2}
\as(\mu^2)\equiv\beta^{(d)}(\as(\mu^2),\epsilon)&=&-\epsilon\as(\mu^2) +
\beta(\as(\mu^2))\nonumber\\
&=&-\epsilon\as(\mu^2) -\beta_0\as^2(\mu^2)+O(\as^3), 
\label{dbetfun}
\eea
in terms of the four-dimensional beta function of
Eq.~(\ref{bet4}), and 
exploiting the $\mu$-independence of $\az$, Eq.~(\ref{alphabare}) gives
\beq
Z^{(\as)}(\as,\epsilon)=\exp\int_0^\alpha \frac{d\alpha'}{\alpha'}\,
\frac{\beta(\alpha')}{\epsilon\alpha'-\beta(\alpha')}
=\left(1+\frac{\beta_0\as}{\epsilon}\right)^{-1}+
O\left(\frac{\as^{k+1}}{\epsilon^k}\right). 
\label{explzet}
\eeq
The renormalized anomalous dimension is found by rewriting Eq.~(\ref{ldef}) as
\beq                                                      
\gamma\left(N,Q^2/\mu^2,\as(\mu^2),\epsilon\right)=       
\sum_{i=1}^{\infty}\sum_{j=0}^i \gamma_{ij}(\epsilon) \,  
\left[\bar\alpha_0(1,\as(Q^2),\epsilon)\right]^{i-j}\,    
\left[\bar\alpha_0(1,\as(Q^2/N^a),\epsilon)\right]^j,
\label{ldef2}
\eeq
and then re-expanding in powers of the renormalized coupling:
\beq                                                      
\gamma\left(N,Q^2/\mu^2,\as(\mu^2),\epsilon\right)=       
\sum_{m=1}^{\infty}\sum_{n=0}^m \gamma^R_{mn}(\epsilon) \,  
\as^{m-n}(Q^2)\,\as^n(Q^2/N^a).
\label{ldef3}
\eeq
The physical anomalous dimension is thus seen to be 
a power series in  $\as(Q^2)$ and $\as(Q^2/N^a)$, with
coefficients $\gamma^R_{mn}$
determined by $\gamma_{ij}(\epsilon)$ and
the renormalization constant $Z^{(\as)}$.  If $\as(Q^2/N^a)$ is
re-expanded in terms of $\as(Q^2)$, the physical anomalous dimension
is seen to be of order $\as(Q^2)$ (and thus vanish when
$\as(Q^2)\to0$), as it ought to.

However, we cannot yet conclude that the four-dimensional physical
anomalous dimension admits an expansion of the form of
Eq.~(\ref{ldef3}), because the coefficients $\gamma^R_{mn}(\epsilon)$ are not
necessarily finite as $\epsilon\to0$. In order to understand this,
it is convenient to separate off the $N$--independent terms in the
renormalized anomalous dimension, \ie\ the terms with
$n=0$ in the internal sum in Eq.~(\ref{ldef3}). Namely, we write
\beq
\gamma\left(N,Q^2/\mu^2,\as(\mu^2),\epsilon\right)=
\hat\gamma^{(l)}\left(\as(Q^2),\as(Q^2/N^a),\epsilon\right)+
\hat\gamma^{(c)}\left(\as(Q^2),\epsilon\right),
\label{gamh}
\eeq
where we have defined
\bea
\hat\gamma^{(l)}\left(\as(Q^2),\as(Q^2/N^a),\epsilon\right)&\equiv&
\sum_{i=1}^{\infty}\sum_{j=1}^i \gamma_{ij}(\epsilon)
\left[\bar\alpha_0(1,\as(Q^2),\epsilon)\right]^{i-j} 
\left[\bar\alpha_0(1,\as(Q^2/N^a),\epsilon)\right]^j
\nonumber\\
&=&\sum_{m=0}^{\infty}\sum_{n=1}^\infty \gamma^R_{m+n\,n}(\epsilon) \,  
\as^m(Q^2)\,\as^n(Q^2/N^a),
\label{gamhl}
\eea
and
\beq
\hat\gamma^{(c)}\left(\as(Q^2),\epsilon\right)\equiv
\sum_{i=1}^{\infty}\gamma_{i0}(\epsilon) \,
\left[\bar\alpha_0(1,\as(Q^2),\epsilon)\right]^i
=\sum_{m=1}^{\infty}\gamma^R_{m0}(\epsilon) \, \as^m(Q^2).
\label{gamhc}
\eeq

Whereas
$\gamma\left(N,Q^2/\mu^2,\as,\epsilon\right)$ is finite in the limit
$\epsilon\to0$, where it coincides with the 
four-dimensional physical anomalous
dimension of Eq.~(\ref{physad}),
$\hat\gamma^{(l)}$ and $\hat\gamma^{(c)}$ are not necessarily
separately finite as $\epsilon\to0$. However, Eq.~(\ref{gamh})
implies that
$\hat\gamma^{(l)}$ and $\hat\gamma^{(c)}$ can be made
finite by adding and subtracting to them a suitable counterterm
$Z^{(\gamma)}\left(\as(Q^2),\epsilon\right)$:
\bea
\gamma^{(l)}\left(\as(Q^2),\as(Q^2/N^a),\epsilon\right)&=&
\hat\gamma^{(l)}\left(\as(Q^2),\as(Q^2/N^a),\epsilon\right)-
Z^{(\gamma)}\left(\as(Q^2),\epsilon\right),
\label{gamlz}
\\
\gamma^{(c)}\left(\as(Q^2),\epsilon\right)&=&
\hat\gamma^{(c)}\left(\as(Q^2),\epsilon\right)+
Z^{(\gamma)}\left(\as(Q^2),\epsilon\right)
\label{gamcz}
\eea
where the functions $\gamma^{(l)}$ and $\gamma^{(c)}$ have a finite
$\epsilon\to 0$ limit. The counterterm cannot depend on $N$ because
of the $N$-independence of $\hat\gamma^{(c)}$.
A possible choice is
therefore
\beq
Z^{(\gamma)}\left(\as(Q^2),\epsilon\right)\equiv
\hat\gamma^{(l)}\left(\as(Q^2),\as(Q^2),\epsilon\right).
\label{zfromgam}
\eeq
With this choice, $\gamma^{(l)}$ is clearly finite for $N=1$, where it
vanishes; because of the $N$-independence of the counterterm, $\gamma^{(l)}$
will then be finite for any $N$. Other forms of the counterterm can
differ from this only in the choice of the finite part, \ie, can be
obtained by adding
to $Z^{(\gamma)}$ Eq.~(\ref{zfromgam}) a
finite function of $\as(Q^2)$. Therefore, they correspond to the
possibility of reshuffling finite $N$--independent terms between
$\gamma^{(l)}$ and $\gamma^{(c)}$. The choice Eq.~(\ref{zfromgam}) is
characterized by the fact that with this choice $\gamma^{(l)}$ is `purely
logarithmic', in that it vanishes at $N=1$ where $\ln N=0$.
We thus get
\bea
\gamma^{(l)}\left(\as(Q^2),\as(Q^2/N^a),\epsilon\right)&=&
\hat\gamma^{(l)}\left(\as(Q^2),\as(Q^2/N^a),\epsilon\right)-
\hat\gamma^{(l)}\left(\as(Q^2),\as(Q^2),\epsilon\right)
\label{gaml}
\\
\gamma^{(c)}\left(\as(Q^2),\epsilon\right)&=&
\hat\gamma^{(c)}\left(\as(Q^2),\epsilon\right)+
\hat\gamma^{(l)}\left(\as(Q^2),\as(Q^2),\epsilon\right),
\label{gamc}
\eea
and the physical anomalous dimension is
\bea
\gamma\left(N,Q^2/\mu^2,\as(\mu^2),\epsilon\right)&=&
\gamma^{(l)}\left(\as(Q^2),\as(Q^2/N^a),\epsilon\right)+
\gamma^{(c)}(\as(Q^2),\epsilon)+O(1/N)\nonumber\\
&=&
\gamma^{(l)}\left(\as(Q^2),\as(Q^2/N^a),\epsilon\right)+O(N^0),
\label{finres}
\eea
where now both $\gamma^{(l)}$ and
$\gamma^{(c)}$ are finite as $\epsilon\to 0$, and 
$\gamma^{(l)}$ provides an
expression of the resummed physical anomalous dimension in the large
$N$ limit, up to non-logarithmic terms.

It is apparent from Eq.~(\ref{gamc}) and the definition
Eq.~(\ref{gamhc})
that $\gamma^{(c)}$ is a power series in $\as(Q^2)$
with finite coefficients. In order to understand  the perturbative
structure of $\gamma^{(l)}$ as well, we notice that
\beq
\gamma^{(l)}\left(\as(Q^2),\as(Q^2/N^a),\epsilon\right)=
\int_1^{N^a}\frac{dn}{n}\, g\left(\as(Q^2),\as(Q^2/n),\epsilon\right)
\label{gamlint}
\eeq
where
\bea
 g\left(\as(Q^2),\as(\mu^2),\epsilon\right)&\equiv&
-\mu^2\frac{\partial}{\partial\mu^2}
\hat\gamma^{(l)}\left(\as(Q^2),\as(\mu^2),\epsilon\right)\nonumber\\ 
&=&
-\beta^{(d)}(\as(\mu^2),\epsilon)\,
\frac{\partial}{\partial\as(\mu^2)}
\hat\gamma^{(l)}\left(\as(Q^2),\as(\mu^2),\epsilon\right),\label{ghdef}
\eea
where $\beta^{(d)}(\as)$ is the $d$-dimensional beta function of
Eq.~(\ref{dbetfun}).  It immediately follows from
Eqs.~(\ref{gamhl}-\ref{ghdef}) that
$g$ is a power series in $\as(Q^2)$ and $\as(\mu^2)$:
\beq
g\left(\as(Q^2),\as(\mu^2),\epsilon\right)=\sum_{m=0}^\infty
\sum_{n=1}^\infty g_{mn}(\epsilon)\,\as^m(Q^2)\,\as^n(\mu^2). 
\label{ghatexp}
\eeq
The sum over powers of $\as(\mu^2)$ in Eq.~(\ref{ghatexp})
starts at $n=1$ because the expansion of  
$\gamma^{(l)}$ contains at least one power of $\as(Q^2/N)$.
The perturbative structure of $\gamma^{(l)}$ is thus dictated by
Eqs.~(\ref{gamlint},\ref{ghatexp}). Because of
Eq.~(\ref{finres}), this provides us with  the desired 
all-order generalization of  
the next-to-leading log resummation formula Eq.~(\ref{ct4})
discussed in Sect.~\ref{NLLres}. 

Our final result for the four-dimensional resummed physical anomalous
dimension is thus
\bea
&&\gamma^{\rm res}(N,\as(k^2))=\int_1^{N^a} \frac{dn}{n}
g(\as(k^2),\as(k^2/n))+O(N^0)
\label{finresad}\\
&&\qquad g(\as(k^2),\as(k^2/n))=\sum_{m=0}^\infty \sum_{n=1}^\infty
g_{mn} \as^m(k^2)\as^n(k^2/n);\quad g_{11}=0.
\label{fing}
\eea
Exploiting the all-order relation between leading $\ln\frac{1}{N}$ and
leading $\ln(1-x)$ resummation derived in Sect.~\ref{Nx},
Eqs.~(\ref{invthirdres},\ref{invfrel}), this result can be
equivalently cast as the Mellin transform of a resummed splitting
function $P^{\rm res}(x,\as(k^2))$:
\bea
&&\gamma^{\rm res}(N,\as(k^2))=\int_0^{1}\!dx x^{N-1}
P^{\rm res}(x,\as(k^2)) +O(N^0)
\label{finresspf}\\
&&\qquad P^{\rm res}(x,\as(k^2))\equiv
\left[\frac{\hat g(\as(k^2),\as(k^2(1-x)^a))}{1-x}\right]_+\label{resspf}\\
&&\qquad \qquad \hat g(\as(k^2),\as(\mu^2))=
-\sum_{p=0}^\infty \frac{\Delta^{(p)}(1)a^p}{p!}
\frac{d^p}{d\ln^p\mu^2}\,g(\as(k^2),\as(\mu^2)).
\label{finghat}
\eea

These resummations generalize to all orders the next-to-leading log
results Eq.~(\ref{ct4}) and Eq.~(\ref{gamexp}), respectively.
The two  forms Eq.~(\ref{finresad}) and~(\ref{finresspf}) 
of the resummation are equivalent up to non-logarithmic
terms, and can be used to compute the resummed evolution factor
Eq.~(\ref{pert2}) $K^{\rm res}(N;Q_0^2,Q^2)$:
\beq
K(N;Q_0^2,Q^2)=\exp\int_{Q_0^2}^{Q^2}\frac{dk^2}{k^2}
\gamma^{\rm res}(N,\as(k^2)).  
\label{finresk}
\eeq
Resummed expressions for the coefficient function and anomalous
dimension in any factorization scheme can be obtained from
Eq.~(\ref{finresk}), as we will discuss in Sect~\ref{schemes}.

\subsection{Next-to-leading resummation}
\label{nlores}

Let us now study  our all-order generalized resummation
at the leading and next-to-leading level. This will also help us to
understand the relation of our resummed result to the stronger one
proven in Ref.~\cite{contopa}.
The lowest--order term in Eq.~(\ref{ghatexp}), used in the expression
of the resummed anomalous dimension Eqs.~(\ref{finres}-\ref{gamlint}),
leads to the leading log resummation:
\bea
\gamma^{(l)}_1&=& -{g}_{01} \int_{\as(Q^2)}^{\as(Q^2/N^a)}
\frac{d\alpha}{\beta^{(d)}(\alpha,\epsilon)}\, \alpha
\nonumber\\ 
&=&\frac{{g}_{01}}{\beta_0}\,
\ln\frac{\beta_0\as(Q^2/N^a)+\epsilon}{\beta_0\as(Q^2)+\epsilon}
+O(\as(Q^2/N)),
\label{llres}
\eea
which reduces to the leading-log result Eq.~({\ref{gll}) as
$\epsilon\to 0$.

Recalling that $\hat\gamma^{(l)}_1$ Eq.~(\ref{gaml}) is independent of
$\as(Q^2)$ and is free of constant (\ie\ $\as$--independent) terms,
Equation (\ref{llres}) uniquely determines the leading log
contribution to $\hat\gamma^{(l)}$:
\beq
\hat\gamma^{(l)}_{1}=\frac{{g}_{01}}{\beta_0}
\ln\left(1+\frac{\beta_0\as(Q^2/N^a)}{\epsilon}
\right),
\label{llsing}
\eeq
This shows explicitly that $\hat\gamma^{(l)}$  
is an analytic function of $\as(Q^2/N)$,
but does not have a finite $\epsilon\to0$ limit. By contrast,
$\gamma^{(l)}_{1}$ Eq.~(\ref{llres}) does have a finite
$\epsilon\to0$ limit, but the limit is no longer an analytic function of
both $\as(Q^2/N)$ and $\as(Q^2)$.

Subleading resummed terms are found by integrating according to
Eq.~(\ref{gamlint}) higher order contributions in the expansion
Eq.~(\ref{ghatexp}), and consistently including higher--order
contributions to the beta function. In general, the term $ g_{ij}$
will generate a next$^{i+j-1}$-to-leading order resummation (when the
beta function is included at leading order).  The NLO resummation of
Eq.~(\ref{ct4}) is recovered if the only subleading contribution is
the $ g_{02}$ term in the expansion Eq.~(\ref{ghatexp}), \ie, if $
g_{11}=0$.

As discussed in the introduction, an all-order generalization of the
resummation Eq. (\ref{ct4}) was derived in Ref.~\cite{contopa}. This
resummation has the form of Eqs.~(\ref{finres}-\ref{gamlint}), but
with $g$ a function of $\as(\mu^2)$ only, \ie, with all $ g_{mn}=0$
when $m>0$ in Eq.~(\ref{ghatexp}). In such case, knowledge of the
first $k$ coefficients $ g_{0n}$, $1\le n\le k$ fully determines the
next$^{k-1}$-to-leading log resummation. These coefficients, in turn,
can be determined by performing a next$^{k-1}$-to-leading fixed-order
computation. The more general resummation of Eq.~(\ref{ghatexp}) is
less predictive. Indeed, if all coefficients $ g_{mn}$ are nonzero,
knowledge of the $k(k+1)/2$ coefficients with $m+n\leq k$ is necessary
in order to determine the next$^{k-1}$-to-leading log resummation. In
turn, this requires a next$^{\frac{k(k+1)}{2}-1}$-to-leading fixed-order
computation. So, while the leading--log resummation is determined by
the $O(\as)$ result, an $O(\as^3)$ computation is required in order to
determine both coefficients $ g_{20}$ and $ g_{11}$ which control the
next-to-leading log resummation, and so forth. 

The conditions under which the more restrictive result of Ref.~\cite{contopa}
holds can be understood by comparing to our approach the derivation of
that result. The approach of Ref.~\cite{contopa} is based on assuming 
the validity of a factorization
formula which is more restrictive than the standard collinear
factorization Eq.~(\ref{pert}). This factorization was 
proven for a wide class of processes in Ref.~\cite{fact}, and it 
implies that the perturbative
coefficient Eq.~(\ref{pert}) in the large-$N$ limit
can be factored as:
\beq
C(N,Q^2/\mu^2,\as(\mu^2))
=C^{(l)}(Q^2/(\mu^2N^a),\as(\mu^2))\,C^{(c)}(Q^2/\mu^2,\as(\mu^2)). 
\label{cfac}
\eeq
or equivalently
\beq
C^{(0)}(N,Q^2,\az)
=C^{(0,l)}(Q^2/N^a,\az,\epsilon)\,C^{(0,c)}(Q^2,\az,\epsilon). 
\label{cfac0}
\eeq
Then, the physical anomalous dimension Eq.~(\ref{gcalc}) manifestly
becomes
\beq
\gamma\left(N,Q^2/\mu^2,\as(\mu^2),\epsilon\right)
=\gamma^{(l)}\left(Q^2/(N^a \mu^2),\as(\mu^2),\epsilon\right)
+\gamma^{(c)}\left(Q^2/\mu^2,\as(\mu^2),\epsilon\right),
\label{gfac}
\eeq
\ie, $\gamma^{(l)}$ depends on
$\as(Q^2/N^a)$ only. Proceeding as above, one then ends up with the
resummation Eqs.~(\ref{finres}-\ref{gamlint}), 
but with $ g= g(\as(\mu^2),\epsilon)$,
or, equivalently, all $ g_{mn}=0$ when $m>0$, as advertised.

By running this argument in reverse, the factorization
Eq.~(\ref{cfac}) is thus seen to be a necessary and sufficient
condition for the validity of the more restrictive factorization
formula of Ref.~\cite{contopa}, where $ g= g(\as(\mu^2))$. This
factorization is in general rather nontrivial. Indeed,  
recalling the expansion Eqs.~(\ref{Cbare},\ref{salame}),
Eq.~(\ref{cfac0}) is seen to be satisfied if and only if
in the large-$N$ limit the coefficients
$C^{(0)}_{n}(\epsilon)$ can be written as
\beq
\label{cfacn}
C^{(0)}_{n}(\epsilon)=\sum_{k=0}^n
F_k(\epsilon)\,G_{n-k}(\epsilon)\,(Q^2)^{-(n-k)\epsilon}
\left(\frac{Q^2}{N^a}\right)^{-k\epsilon}.
\eeq
Whereas an
all--order proof of Eq.~(\ref{cfacn}) is highly nontrivial, the
structure of our more general, but weaker resummation
Eqs.~(\ref{finres}-\ref{gamlint}) implies that a fixed order proof of
the factorization is in fact sufficient for the corresponding resummed
result to hold. 

Consider the simplest nontrivial case, namely,
the next-to-leading log resummation: this is determined by knowledge
of the coefficients ${g}_{20}$ and ${g}_{11}$. The stronger
resummation holds when ${g}_{11}=0$. But in order to check whether
this is the case or not it is sufficient to verify that the desired
factorization Eq.~(\ref{cfacn}) holds to lowest nontrivial order,
namely that
\beq
C^{(0)}_2(N,Q^2,\epsilon)=G_2(\epsilon)(Q^2)^{-2\epsilon}
+F_1(\epsilon)G_1(\epsilon)(Q^2)^{-\epsilon}
\left(\frac{Q^2}{N^a}\right)^{-\epsilon} +F_2(\epsilon)
\left(\frac{Q^2}{N^a}\right)^{-2\epsilon}+O(1/N).
\label{strongfact}
\eeq

Equivalently, we can compute the coefficients $g_{01}$, ${g}_{20}$ and
${g}_{11}$ which determine the next-to-leading log resummation by
comparison with fixed-order results: in order to calculate three
coefficients, we need a next-to-next-to leading fixed order
computation. 
Indeed,
expanding Eqs.~(\ref{finresad}) 
in powers of $\as(Q^2)$ to order $\as^3$ we get
\bea
\label{gammaf2r}
&&\gamma(N,\as(Q^2))\\
&&=
-g_{01} \as(Q^2)\ln\frac{1}{N}
+\left[\frac{\beta_0 g_{01}}{2}\ln^2\frac{1}{N}
-(g_{11}+g_{02})\ln\frac{1}{N}\right] \as^2(Q^2)
\nonumber\\
&&-\left[\frac{\beta_0^2 g_{01}}{3} \ln^3\frac{1}{N}
-\left(\frac{\beta_0 g_{11}}{2}+\beta_0 g_{02}+\frac{\beta_1 g_{01}}{2}\right)
\ln^2\frac{1}{N}
+\left(\frac{\beta_1g_{02}}{\beta_0}-\frac{\beta_1^2g_{01}}{\beta_0^2} \right)
\ln\frac{1}{N}\right]\as^3(Q^2)\nonumber\\
&&+O(\as^4)+O(N^0).
\nonumber
\eea
Hence, after having determined $g_{01}$ from the leading log result,
the $O(\as^2)$, $O(\ln N)$ term only determines the
combination $g_{11}+g_{02}$, but the $O(\as^3)$, $O(\ln^2 N)$ provides
an independent linear combination. 

The physical anomalous dimension is in turn determined by the
coefficient function $C$ Eq.~(\ref{pert}) and the standard
anomalous dimension $\gamma^{AP}$ 
Eq.~(\ref{evolF}) according to Eq.~(\ref{generic}), which through
order $\as^3$ gives
\beq
\label{gammaf2p}
\gamma(N,\as(Q^2))=
\gamma_1^{\rm AP}\as(Q^2)
+(\gamma_2^{\rm AP}-\beta_0 C_1)\as^2(Q^2)
+(\gamma_3^{\rm AP}-2\beta_0 C_2+\beta_0 C_1^2-\beta_1 C_1)\as^3(Q^2)
+O(\as^4),
\eeq
where  $\gamma_i^{\rm AP}$  and $C_i$ are the order-$\as^i$
coefficients in the expansion of $\gamma^{AP}$ and $C$  respectively. 

The coefficients  $C$ are known up to $O(\as^3)$~\cite{cfp}
for Drell-Yan~\cite{dynnlo} and deep-inelastic~\cite{disnnlo}
processes, while 
$\gamma^{AP}$ is known (in the \MS\ scheme) up to
$O(\as^2)$~\cite{cfp}, and has been very recently determined
to $O(\as^3)$ in the large $N$
limit~\cite{scia}.  However, knowledge of the $O(\as^2)$ anomalous
dimension  is in fact sufficient to determine $g_{11}$, 
thanks to the fact that in the \MS\ scheme the anomalous
dimension is linear in $\ln N$ to all orders~\cite{kor}, so its
knowledge to $O(\as^2)$ is sufficient to establish the vanishing of $g_{11}$.

To see how this works, consider explicitly the case
of the deep-inelastic structure function $F_2$. The coefficients in
the \MS\ scheme are
\bea
&&C^{q}_1=\frac{C_F}{4\pi} \left[4I_1-3I_0-4\zeta(2)-9\right] +O(1/N)
\label{cqone}\\
&&C^{q}_2=\frac{1}{8\pi^2}\Bigg\{
  C_F^2 \left[4I_3-9I_2-\left(16 \zeta(2)+\frac{27}{2}\right) I_1
    +\left(\frac{51}{4}+18\zeta(2)-4 \zeta(3)\right) I_0\right]
\nonumber\\
&&\phantom{\frac{1}{8\pi^2}\Big\{}
+C_A C_F \left[-\frac{11}{3}I_2
        +\left(\frac{367}{18}-4 \zeta(2)\right) I_1
        +\left(-\frac{3155}{108}+\frac{22}{3} \zeta(2)+20 \zeta(3)\right) 
I_0\right]
\nonumber\\
&&\phantom{\frac{1}{8\pi^2}\Big\{}
      +n_f C_F \left[\frac{2}{3} I_2-\frac{29}{9} I_1
+\left(\frac{247}{54}-\frac{4}{3} \zeta(2)\right) I_0\right]\Bigg\}+O(N^0),
\label{cqdis}\eea
where $\zeta$ is the Riemann $\zeta$ function, $C_F$ and $C_A$ are the
usual quadratic Casimir operators, and
$I_p$ have been defined in Eq.~(\ref{ipdef}) and, as proven in
Sect.~\ref{Nx}, contain the logarithmic $N$ dependence. Note that  
$O(N^0)$ terms have been included in Eq.~(\ref{cqone}) in view of the fact
that $C_1^2$  appears
in Eq.~(\ref{gammaf2p}). The gluon coefficient to next-to-leading
order is $O(1/N)$. 
The 
\MS\ anomalous dimension in turn is
\bea
&&{\gamma_1^{qq}}^{\rm AP}=\frac{C_F}{\pi}\, I_0 +O(N^0)
\\
&&{\gamma_2^{qq}}^{\rm AP}=
\frac{1}{4\pi^2}\left[-\frac{10}{9} n_f C_F
+\left(\frac{67}{9}-2\zeta(2)\right) C_A C_F\right]\, I_0+O(N^0),
\label{gamqq}\eea
while $\gamma^{qg}$ and $\gamma^{gq}$ are $O(1/N)$. The gluon
anomalous dimension $\gamma^{gg}$
does contain soft logs, but its  contribution to $F_2$ only goes
through via interference with $C_g$, and it is thus power suppressed.
Therefore, we see that through next-to-leading order there is no
operator mixing, and only the quark sector contributes to the large
$N$ limit.
 Substituting in Eq.~(\ref{gammaf2p}), and solving for $g_{11}$ and $g_{02}$
we get, after tedious but straightforward algebra 
\bea
&&g_{01}=-\frac{C_F}{\pi}
\\
&&g_{02}=\frac{1}{144\pi^2}
\left[(24\gamma_E+58)C_F n_f+(12\pi^2-132\gamma_E-367)C_AC_F\right]
\\
&&g_{11}=0.\eea
In Drell-Yan there is likewise no mixing, and the check that 
$g_{11}=0$, performed  in Ref.~\cite{dyres}, 
follows the same lines.

Hence, by combining available fixed-order results with our general
resummation Eq.~(\ref{finresad}), we reproduce the
next-to-leading log resummation Eq.~(\ref{ct4}). 
Beyond next-to-leading order the general result 
Eq.~(\ref{finresad}) holds, and
will have to be used for the full matrix of
two-by-two resummed physical anomalous dimensions. 

\sect{Momentum-space resummation}

We have seen that resummed results can be compactly expressed in terms
of the physical anomalous dimension Eq.~(\ref{finresad}), or in terms
of a resummed splitting function Eq.~(\ref{resspf}) obtained from it
by inverse Mellin transformation, up to non-logarithmic terms. In the
former case, the resummation is cast in Mellin $N$ space, and one
resums large $\ln \frac{1}{N}$ (at large $N\to\infty$). In
the latter case, the splitting function is defined in momentum space,
\ie, as a function of the momentum fraction $x$ and one resums large $\ln
(1-x)$ (at large $x\to1$).  In this section we address the question
whether it is possible to construct a resummation of $\ln(1-x)$
directly at the level of the momentum-space cross section
$\sigma(x,Q^2)$. 
Indeed, even when the resummation is performed at the level of splitting
functions, the cross section is written as the inverse Mellin
transform of the exponentiated result Eq.~(\ref{gamexp}).  It would
seem more natural if the momentum-space cross section $\sigma(x,Q^2)$
could be written directly as the exponential of a suitable leading,
next-to-leading,\dots~$\ln(1-x)$ function.

A general way of performing the inverse Mellin transform of a
leading-, next-to-leading, \dots, $\ln\frac{1}{N}$ function up to terms
which are suppressed by powers of $1-x$ was given in Sect.~\ref{Nx},
Eq.~(\ref{lxtolnmel}). Using this result, it is easy to rewrite the
$N$-space resummation factor Eq.~(\ref{finresk}) as the Mellin
transform of an $x$ space resummed expression:
\bea
&&K^{\rm res}(N;Q_0^2,Q^2)=\exp 
\int_{Q_0^2}^{Q^2}\frac{dk^2}{k^2}\,\gamma^{(l)}
(\as(k^2),\as(k^2/N^a))+O(N^0)\label{melxres}\\
&&\quad\quad
=1+\sum_{k=1}^\infty \frac{1}{k!} 
\int_0^1\!dx\,x^{N-1}\sum_{n=1}^\infty \frac{\Delta^{(n-1)}(1)}{(n-1)!}
\nonumber\\
&&\quad\quad\times
\left[\frac{1}{1-x}\frac{d^n}{d\ln^n(1-x)}
\left(\int_{Q_0^2}^{Q^2}\frac{dk^2}{k^2}\,\gamma^{(l)}
(\as(k^2),\as(k^2(1-x)^a))\right)^k\right]_++O(N^0),\nonumber
\eea
where $\gamma^{(l)}$  is the resummed physical anomalous dimension
Eq.~(\ref{gamlint}), and, even though the resummation factor $K^{\rm res}$
is defined up to non-logarithmic terms, we have normalized it by the
requirement that $K^{\rm res}\to1$ as $\as\to0$.  
This immediately leads to the identification of the momentum-space
resummation factor:
\bea
&&K^{\rm res}(x;Q_0^2,Q^2)=\delta(1-x)-\>\sum_{k=1}^\infty \frac{1}{k!}
\sum_{n=1}^\infty \frac{\Delta^{(n-1)}(1)}{(n-1)!}
\nonumber\\
&&\quad\quad\quad\times
\left[\frac{d}{dx}\frac{d^{n-1}}{d\ln^{n-1}(1-x)}
\left(\int_{Q_0^2}^{Q^2}\frac{dk^2}{k^2}\,\gamma^{(l)}
(\as(k^2),\as(k^2(1-x)^a))\right)^k\right]_+.
\label{xres}
\eea

We can now  derive a closed form of the sum of the exponential
expansion in Eq.~(\ref{xres}), using a representation of the $+$
distribution given in Ref.~\cite{paolo}: 
\bea
&&K^{\rm res}(x;Q_0^2,Q^2)=-\lim_{\eta\to0^+}\sum_{n=1}^\infty
\>\frac{\Delta^{(n-1)}(1)}{(n-1)!}
\frac{d^{n-1}}{d\ln^{n-1}(1-x)}
 \nonumber\\
&&\quad\quad\quad
\frac{d}{dx}\left[\theta(1-x-\eta)
\exp\int_{Q_0^2}^{Q^2}\frac{dk^2}{k^2}\,\gamma^{(l)}
(\as(k^2),\as(k^2(1-x)^a))\right]
\label{formsum}
\eea
where $\theta$ is the Heaviside function. The equality holds in
the sense of distributions: it is proven by folding $K^{\rm res}$
with a test function, integrating by parts, and noting that
$\theta(-\eta)=0$ and $\gamma^{(l)}(\as(Q^2),\as(Q^2))=0$
because of Eq.~(\ref{gaml}).  It is also convenient to define a
momentum-space resummation exponent $E^{\rm res}(x;Q_0^2,Q^2)$ through
the implicit equation
\beq
K^{\rm res}(x;Q_0^2,Q^2)=-\lim_{\eta\to0^+}\frac{d}{dx}\left[\theta(1-x-\eta)
\exp E^{\rm res}(x;Q_0^2,Q^2)\right].
\label{resedef}
\eeq

Terms with $n=1,2,\dots$ in Eqs.~(\ref{xres},\ref{formsum}) 
generate leading, 
next-to-leading,\dots $\ln(1-x)$ contributions to the resummation
factor. Hence, if $\gamma^{(l)}$ is computed at the leading log level
Eq.~(\ref{gll}),
only the  $n=1$ terms should be retained, 
\bea
K^{LL}(x;Q_0^2,Q^2)&=&
-\lim_{\eta\to0^+}\frac{d}{dx}\left[\theta(1-x-\eta)
\exp\left(-\int_{Q_0^2}^{Q^2}\frac{dk^2}{k^2}\,
\frac{g_1}{\beta_0}\ln\left(1+\beta_0\as(k^2)\ln(1-x)^a
\right)\right)\right]\nonumber\\
&=&
-\lim_{\eta\to0^+}\frac{d}{dx}\left[\theta(1-x-\eta)
\exp E^{LL}\left(x;Q_0^2,Q^2\right)\right],\label{formllsum}
\eea
where
\beq
E^{LL}\left(x;Q_0^2,Q^2\right)
=-\frac{g_1}{\beta_0}\ln(1-x)^a\left[-\ln\as
+\frac{1+\as\beta_0\ln(1-x)^a}{\as\beta_0\ln(1-x)^a}
\ln\left(1+\as\beta_0\ln(1-x)^a\right) 
\right]^{\as(Q^2)}_{\as(Q_0^2)}.
\label{formllexp}
\eeq
At next-to-leading and higher orders, the action of subsequent terms
in the series of derivatives Eq.~(\ref{formsum})
should be accordingly truncated so that the resummation exponent
$E^{\rm res}$ is computed to the given order.

We would thus be led to conclude that Eq.~(\ref{formllsum}), and its
obvious subleading generalizations, provide us with the desired form
of the leading-, next-to-leading-,\dots,~$\ln(1-x)$ resummation
factor, which can then be used to construct the corresponding resummed
momentum-space cross sections.  Surprisingly, however, it turns out
that the formal expression Eq.~(\ref{formllsum}) is actually
ill-defined: the corresponding resummed cross section diverges. This
difficulty was first  pointed out in the fixed coupling limit
in Ref.~\cite{paolo}.
Also surprisingly, this divergence is a
consequence of the leading- (next-to-leading,\dots)~$\ln(1-x)$
truncation: it is absent if the Mellin transform of the leading-
(next-to-leading,\dots)~$\ln\frac{1}{N}$ result is performed up to
power suppressed terms, \ie, by retaining all terms in the sum over
$n$ Eq.~(\ref{formsum}).

To see this divergence, 
recall that the resummed evolution  factor $K^{\rm res}$ 
according to Eq.~(\ref{pert2}) satisfies
\beq
\sigma(z,Q^2)=\int_z^1\!\frac{dx}{x}\,K^{\rm res}(x;Q_0^2,Q^2)\,
\sigma\left(\frac{z}{x},Q_0^2\right)
\equiv\int_0^1\!dx\,K^{\rm res}(x;Q_0^2,Q^2)\,\tau(x)
\label{divact}
\eeq
where
\beq
\tau(x)=\frac{1}{x}\,\sigma\left(\frac{z}{x},Q_0^2\right)\,\theta(x-z).
\eeq
Integrating by parts and noting that $\tau(0)=0$, we get
\beq
\sigma(z,Q^2)=
\int_0^1\!dx\,\frac{d\tau(x)}{dx}\, \exp E^{\rm res}(x;Q_0^2,Q^2).
\label{divactp}
\eeq
Specializing to the leading log resummation Eq.~(\ref{formllsum}) 
\bea
&&\exp E^{LL}(x;Q_0^2,Q^2)
\label{paofc}
\\&&=\left(\frac{\as(Q^2)}{\as(Q_0^2)}\right)^{G_0
\ln(1-x)^a} \exp\left[G_1
\left(\as(Q^2)-\as(Q_0^2)\right)\ln^2(1-x)^a\right] \exp
O(\as^2),
\nonumber
\eea
where we have expanded out the leading log exponent
Eq.~(\ref{formllexp}) in powers of $\as$, and $O(\as^2)$ denotes the
next-to-next-to leading term in this expansion, which is in fact of
order $(\as^2(Q^2)-\as^2(Q_0^2))\ln^3(1-x)^a=\as^3(Q^2)\ln^3(1-x)^a\left[1+
O(\as(Q^2))\right]$. Explicitly, the coefficients of
the first two orders of this expansion are $G_0=
g_1/\beta_0$ and $G_1=-g_1/2$.  

Now, as $x\to1$, the $O(\as)$
contribution to $E^{LL}$ diverges as $\ln^2(1-x)$, so $\exp E^{LL}$
diverges faster than any power of $1-x$. However, $\tau(x)$ is
manifestly regular as $x\to 1$, so the integral Eq.~(\ref{divactp})
does not exist.  In fact, matching with the leading-order anomalous
dimension shows that $g_1=-C_F/\pi<0$, so the integral diverges for
$Q^2<Q_0^2$; however (regardless of the sign of $ g_1$) the presence
of this divergence means that the perturbative expansion of $\sigma$
in powers of $\as$ has vanishing radius of convergence.

This can also be seen explicitly by using the leading log
$K^{LL}$ Eq.~(\ref{formllsum}) in Eq.~(\ref{divact}):
\bea
\sigma(z,Q^2)&=&\int_0^1\!dx\, \frac{d\tau(x)}{dx}\,
\left(\frac{\as(Q^2)}{\as(Q_0^2)}\right)^{G_0\ln(1-x)^a}
\sum_{p=0}^\infty
\frac{\left[G_1\left(\as(Q^2)-\as(Q_0^2)\right)\right]^p}{p!}
\ln^{2p}(1-x)^a
\nonumber\\ &&\times\exp O(\as^2).
\label{facdivser}
\eea
The generic term in the sum over $p$ is
\bea
K_p&\approx& \frac{d\tau(x)}{dx}
\Bigg|_{x=1}
\frac{\left[a^2 G_1\left(\as(Q^2)-\as(Q_0^2)\right)\right]^p}{p!}\,
C_{2p}\nonumber\\
C_{2p}&\equiv&
\int_0^1\!dx\, \left(1-x\right)^{ a
G_0\ln\frac{\as(Q^2)}{\as(Q_0^2)}}
 \ln^{2p}(1-x) \label{gendtermsum}\\
&=&
 (2p)!\frac{1}{\left(1+a
G_0\ln\frac{\as(Q^2)}{\as(Q_0^2)}\right)^{2p+1}}\nonumber
\eea
where in we have Taylor expanded $\tau(x)$ about $x=1$,
and neglected terms which are suppressed by powers of $1-x$,
consistent with the fact that the resummed distribution $K^{\rm res}$
is determined in the same approximation. The factorial growth of
$C_p$ Eq.~(\ref{gendtermsum}) implies that the sum over $p$ in
Eq.~(\ref{facdivser}) diverges factorially.

Higher order contributions in the expansion of
$K^{LL}$ in  powers of $\as$ will make this divergence worse. Indeed,
the $O(\as^k)$ contribution is proportional to $\ln^k(1-x)$, so the
factorial divergence of the corresponding expansion is more
severe. Furthermore, the expansion of $K^{LL}$ amounts essentially to
the expansion of $\ln(1+\as \beta_0 \ln(1-x)^a)$ in powers of $\as$,
which, since $x<1$ is a series of terms with the same sign, so these
divergences add up. 

In fact, the leading log exponent Eq.~(\ref{formllexp}) blows up when
$\ln(1-x)^a=-\frac{1}{\as\beta_0}$. This divergence corresponds 
to the Landau pole in the coupling constant $\as(k^2(1-x)^a)$
which appears in the $x$--space resummation factor
Eq.~(\ref{xres}). Because of the Landau pole, the expansion of the
resummed exponent 
$E$ in powers of $\as$ has a finite radius of convergence. 
However, as first clarified in Ref.~\cite{paolo},
the factorial divergence Eq.~(\ref{facdivser}) is 
entirely unrelated to the Landau pole. Indeed, we have seen that
it is present even if
we retain only one term in the expansion of the resummed exponent:
rather, this divergence plagues the series expansion of the
exponential, when integrated term by term.

Now, we exploit the power of our all-order relation between leading
$\ln(1-x)$ and leading $\ln\frac{1}{N}$ series to show that this
factorial divergence is just a byproduct of the truncation to
leading (or next-to-leading, etc.) level of the full
expression Eq.~(\ref{xres}) of the $x$--space resummation factor. 

To this purpose, let us again concentrate on the leading log
expression, and separate off the first few terms  in the expansion of
the resummed exponent in powers
of $\as$: retaining
all terms in the sum over $n$ Eq.~(\ref{xres}), we get 
\bea
&&\sigma(z,Q^2)=\int_0^1\!dx \frac{d\tau}{dx}
\sum_{n=1}^\infty \frac{\Delta^{(n-1)}(1)}{(n-1)!}
\frac{d^{n-1}}{d\ln^{n-1}(1-x)}\nonumber\\
&&\qquad\Bigg[
\left(\sum_{p=0}^\infty
\frac{\left[G_0\ln\frac{\as(Q^2)}{\as(Q_0^2)}\ln(1-x)^a\right]^p}
{p!}\right)
\left(\sum_{q=0}^\infty
\frac{\left[ G_1\left(\as(Q^2)-\as(Q_0^2)\right)\right]^q}{q!}
\ln^{2q}(1-x)^a\right)\label{serprod}\\
&&\qquad\qquad\times
\left(\sum_{r=0}^\infty
\frac{\left[G_2\left(\as^2(Q^2)-\as^2(Q_0^2)\right)\right]^r}{r!}
\ln^{3r}(1-x)^a\right)\exp
O(\as^3)\Bigg]\nonumber\\
&&\phantom{\sigma(z,Q^2)}=\int_0^1 \!dx \frac{d\tau}{dx} 
\Bigg[\exp O(\as^3) 
\left(\sum_{r=0}^\infty
\frac{\left[G_2\left(\as^2(Q^2)-\as^2(Q_0^2)\right)\right]^r}{r!}
a^{3r}\frac{d^{3r}}{d\eta^{3r}}\right) 
\nonumber\\
&&\qquad\qquad\qquad \times
\left(\sum_{q=0}^\infty
\frac{\left[ G_1\left(\as(Q^2)-\as(Q_0^2)\right)\right]^q}{q!}
a^{2q}\frac{d^{2q}}{d\eta^{2q}}\right)\label{serder}\\
&&\qquad\qquad\qquad\qquad\times
\left(\sum_{p=0}^\infty
\frac{\left[G_0\ln\frac{\as(Q^2)}{\as(Q_0^2)}\right]^p}
{p!}a^p\frac{d^p}{d\eta^p}\right)
\Bigg]\frac{(1-x)^\eta}{
\Gamma(1+\eta)}\Bigg|_{\eta=0},\nonumber
\eea
where in the last step we have used the definition
Eq.~(\ref{smallgdef}) of $\Delta^{(n)}$, and the limit $\eta\to0$
should be taken after performing all derivatives.

Now, it is easy to compute the inner ``leading order'' sum (index $p$):
$\exp \frac{d}{d\eta}$  acts as a finite translation operator. We get
\bea
&&\sigma(z,Q^2)=\int_0^1\!dx\,\frac{d\tau}{dx}
\Bigg[\exp O\left(\as^3\right) 
\left(\sum_{r=0}^\infty
\frac{\left[G_2\left(\as^2(Q^2)-\as^2(Q_0^2)\right)\right]^r}{r!}
a^{3r}\frac{d^{3r}}{d\eta^{3r}}\right) \nonumber \\
&&\qquad\qquad\qquad \times
\left(\sum_{q=0}^\infty
\frac{\left[ G_1\left(\as(Q^2)-\as(Q_0^2)\right)\right]^q}{q!}
a^{2q}\frac{d^{2q}}{d\eta^{2q}}\right)\Bigg]
\frac{(1-x)^{\eta+a G_0\ln\frac{\as(Q^2)}{\as(Q_0^2)}}}
{\Gamma(1+\eta+a G_0
\ln\left(\frac{\as(Q^2)}{\as(Q_0^2)}\right))}\Bigg|_{\eta=0}. 
\label{sersumzero}
\eea
We see therefore that, due to the inclusion of subleading corrections
of the sum Eq.~(\ref{xres}), the generic term in the sum over $q$ is now
\bea
K_q&\approx&\frac{d\tau(x)}{dx}\Bigg|_{x=1} 
\frac{\left[a^2 G_1\left(\as(Q^2)-\as(Q_0^2)\right)\right]^q}{q!}
\,C_{2q}\nonumber\\
C_{2q}&=& 
\frac{d^{2q}}{d\eta^{2q}} \frac{1}{\Gamma\left(2+\eta+
a G_0\ln\frac{\as(Q^2)}{\as(Q_0^2)}\right)}\Bigg|_{\eta=0}, 
\label{genctermsum}
\eea
where we have again Taylor expanded $\tau(x)$ about $x=1$.

Comparing this with the previous result Eq.~(\ref{gendtermsum}) we see
that the  power in the denominator is replaced by a Gamma function. 
This is enough to tame the factorial divergence as we now show. We
observe that
\beq
\sum_{q=0}^\infty
\frac{C_q}{q!} z^q=
\frac{1}{\Gamma\left(2+z+A_0\ln\frac{\as(Q^2)}{\as(Q_0^2)}\right)}:
\label{tayser}
\eeq
hence, $C_q$ are the coefficients of the Taylor
series expansion of an entire function, which, therefore, has infinite
radius of convergence (note that if we had used the result
Eq.~(\ref{gendtermsum}) we would have found instead a finite radius of
convergence). This in turn implies that
\beq
\lim_{q\to\infty} \,\frac{C_{q+1}}{(q+1)!}\frac{q!}{C_q}
=\lim_{q\to\infty} \,\frac{1}{q+1}\,\frac{C_{q+1}}{C_q}=0,
\label{infconva}
\eeq
and consequently
\beq
\lim_{q\to\infty} \,\frac{C_{2q+1}}{(q+1)!}\frac{q!}{C_{2q}}
=\lim_{q\to\infty} \frac{1}{q+1}\,\frac{C_{2q+1}}{C_{2q}}
=0.
\label{infconvb}
\eeq
This means that the coefficients $C_{2q}$ no longer diverge
factorially: in fact, they are also coefficients of the Taylor
series expansion of an entire function, with an infinite convergence
radius.

 Proceeding in
the same way, we can then show that the sum over $r$ in
Eq.~(\ref{serder}), \ie, the expansion of the $\exp O(\as^2)$ 
contribution,
 is also convergent, and so forth.  Therefore, we conclude that order
by order in perturbation theory, the resummed result Eq.~(\ref{xres})
leads to a convergent perturbative expansion, but if this is turned
into a leading-, next-to-leading,\dots $\ln(1-x)$, a meaningless
divergent expansion is found. 

In summary, we have seen that the leading-, next-to-leading,\dots
$\ln(1-x)$ expressions of the momentum-space cross section
$\sigma(x,Q^2)$ are all ill-defined. However, if the resummation is
formulated at the leading-, next-to-leading,\dots
$\ln\frac{1}{N}$ level, and the resummed $\sigma(x,Q^2)$ is defined as
the inverse Mellin of the ensuing result, up to power corrections,
then finite results are obtained. 

\sect{Factorization schemes and matching}
\label{schemes}

The resummed results derived in Sect.~\ref{RGI} provide us with a
prediction for the resummed scale dependence of the physical cross
section, expressed by the physical anomalous dimension $\gamma$. In
practice, we would like to have resummed predictions for the
coefficient function $C$ Eq.~(\ref{pert}) and standard anomalous
dimension $\gamma^{\rm AP}$ Eq.~(\ref{evolF}). Because of
Eq.~(\ref{generic}), that relates $\gamma$ to $\gamma^{\rm AP}$ and
$C$, all the relevant information is contained in the physical
anomalous dimension, and the resummed anomalous dimension and
coefficient function can be determined from it. In order to do this,
we must suitably combine resummed and unresummed results, and
choose a factorization scheme.

The simplest choice consists of choosing a physical
factorization scheme, defined by
\beq
\sigma(N,Q^2)=F(N,Q^2)
\label{physscheme}
\eeq
so that
\beq
C(N,1,\as(Q^2))=1;\quad
\gamma^{\rm AP}(N,\as(Q^2))=\gamma(N,\as(Q^2)). 
\label{physschemebis}
\eeq
%In this scheme, 
%\beq
%C(N,Q^2/\mu^2,\as(\mu^2))=\exp\int_{\mu^2}^{Q^2}\frac{dk^2}{k^2}
%\,\gamma^{\rm AP}(N,\as(k^2))
%\label{physschemecf}
%\eeq
%so that
%\beq
%\sigma(N,Q^2)=C(N,Q^2/\mu^2,\as(\mu^2))\,F(N,\mu^2).
%\eeq
Since now the physical and standard anomalous dimension coincide,
Eq.~(\ref{finresad}) gives us a resummed expression of this anomalous
dimension, up to non-logarithmic terms. 
A full expression of the
anomalous dimension, valid both at large and small $N$, 
can be obtained by combining these resummed results with standard
fixed-order ones, and subtracting the double counting. Specifically, a
next-to-leading order and next-to-leading log expression of the
physical anomalous dimension is given by
\bea
\gamma(N,\as(k^2))&=&\as(k^2) \gamma^{(0)}(N)+\as^2(k^2) \gamma^{(1)}(N)+
\int_1^{N^a}\frac{dn}{n} \left[g_{01} \as(k^2/n)+g_{02}
\as^2(k^2/n)\right] 
\nonumber\\
&&-\left[-g_{01}\as(k^2) \ln\frac{1}{N^a}
+\as^2(k^2)\left(\frac{g_{01} \beta_0}{2} \ln^2\frac{1}{N^a} 
-g_{02}\ln\frac{1}{N^a}\right)\right],
\label{nlopadres}
\eea
where $\gamma^{(0)}$ is the standard leading-order anomalous
dimension, $\gamma^{(1)}$ is the two-loop physical anomalous dimension,
obtained using Eq.~(\ref{generic}) from the next-to-leading order
anomalous dimension and one-loop coefficient function, and the 
term in square brackets
is the double-counting subtraction of the logarithmic
contributions to $\gamma^{(0)}$ and $\gamma^{(1)}$.

A different resummed expression can be obtained by using the
expression Eq.~(\ref{finresspf}) of the resummed
physical anomalous dimension in terms of the resummed splitting
function Eq.~(\ref{finresad}). In this case, Eq.~(\ref{nlopadres}) is
replaced by
\bea
\gamma(N,\as(k^2))&=&\int_0^1\!dx\, x^{N-1}
\Bigg\{\as(k^2) P^{(0)}(x)+\as^2(k^2) P^{(1)}(x)
\nonumber\\
&&
+\left[\frac{\hat g_{01} \as(k^2(1-x)^a)+\hat g_{02} \as^2(k^2(1-x)^a)}{1-x}
\right]_+
\nonumber\\
&&-\left[\frac{ \hat g_{01}\as(k^2)
                +\as^2(k^2)
                \left(\hat g_{02}-\hat g_{01}\beta_0\ln(1-x)^a\right)
                }{1-x}     \right]_+
\Bigg\},
\label{nlopsfres}
\eea
where $P^{(i)}(x)$ are defined through
\beq
\gamma^{(i)}(N,\as)=\int_0^1\!dx\, x^{N-1} P^{(i)}(x),
\label{gamfromp}
\eeq
and the relation between $g$ and $\hat g$ was given in
Eq.~(\ref{finghat}) (or in Eq.~(\ref{coefrel}) to next-to-leading
order). The two forms of the resummed anomalous dimension
Eq.~(\ref{nlopadres}) and Eq.~(\ref{nlopsfres}) coincide at order
$\as^2$ and at the next-to-leading log level, but differ by
next-to-next-to leading terms, as well as by $O(1/N)$ terms which are
at least $O(\as^3)$. Hence, their difference might be used to estimate
the impact of the next-to-next-to-leading resummation in a
factorization-scheme independent way.

Resummed coefficient function and anomalous dimension in a generic
scheme can be obtained by generalizing the conventional factorization
scheme change at the resummed level.
A scheme change is performed
by redefining the parton distribution by a finite function $Z(N,\as)$,
which can be viewed as a redefinition of the finite part of the
multiplicative renormalization $Z^{(C)}$ of the hard coefficient,
Eq.~(\ref{mulnor})
\beq
F'(N,\mu^2)= Z(N,\as(\mu^2)) F(N,\mu^2).\label{schch}
\eeq
Upon the scheme change, the anomalous dimension and coefficient
function change by
\bea
&&\ln C'(N,Q/\mu^2,\as(\mu^2) )= \ln C(N,Q/\mu^2,\as(\mu^2) ) -
\ln Z(N,\as(\mu^2))\label{cch}\\
&&{\gamma'}^{\rm AP}(N,\as(\mu^2))= 
\gamma^{\rm AP}(N,\as(\mu^2))+\mu^2 \frac{d}{d\mu^2} \ln Z(N,\as(\mu^2)). 
\label{gamch}
\eea

A conventional (unresummed) scheme change is performed by computing
$Z$ to fixed order in $\alpha_s$:
\beq
Z^{\rm u}(N,\as(\mu^2))= 1+Z_1(N) \as(\mu^2)+
Z_2(N)\as^2(\mu^2)+\dots\,.\label{zexp}
\eeq
The term $Z_k$ affects   the
next$^k$-to-leading and higher order contributions to the anomalous
dimension. 
At the resummed next$^k$-to-leading log
level, the physical anomalous dimension is computed
including terms of $O(\as^{n+k} \ln^n N)$ to all orders in $\as$. Such
terms are reshuffled between the coefficient function and the parton
distribution by performing a scheme change Eq.~(\ref{schch}) with 
\bea
&&Z^{\rm res}(N,\as)=\exp\left[\frac{1}{\as} z_{0}\left(\as
\ln\frac{1}{N^a}\right)+z_{1}\left(\as
\ln\frac{1}{N^a}\right)+\as z_{2}\left(\as
\ln\frac{1}{N^a}\right)+\dots\right]\label{zresdef}\\
&&\qquad z_{k}=\sum_{i=2}^\infty z_{k}^{(i)} \left(\as
\ln\frac{1}{N^a}\right)^i,\label{zterm}
\eea
where $z_{k}$ affects the next$^k$-to-leading log and higher
contributions to the
resummed anomalous dimension $\gamma^{\rm AP}$ and coefficient
function $C$. 

Even though any resummed scheme change of the form Eq.~(\ref{zresdef})
is in principle acceptable (\ie,  it can be obtained by redefining the finite
part of $Z^{(C)}$ Eq.~(\ref{mulnor})), in practice it is convenient to
consider in particular resummed scheme changes which preserve the
structure Eq.~(\ref{finresad})  of the resummed anomalous dimension,
\ie\ such that after scheme change Eq.~(\ref{gamch}) the anomalous
dimension can still be written in the form of Eq.~(\ref{finresad}),
and the scheme change only affects the numerical values of the
coefficients $g_{ij}$. Such scheme changes are performed by choosing
$Z$ to be of the form
\bea
&&Z^{\rm res}(N,\as(\mu^2))
=\exp\int_1^{N^a} \frac{dn}{n}
\int_{\Lambda^2}^{\mu^2}\frac{dk^2}{k^2} \left(z(\as(k^2),\as(k^2/n))-
z(\as(k^2),\as(k^2))\right)
\label{zrresdef}\\
&&\qquad z(\as(\mu^2),\as(\mu^2/n))=\sum_{i=0}^\infty\sum_{j=1}^\infty 
z_{ij}\, \as^i(\mu^2)\,\as^j(\mu^2/n);\quad z_{11}=0,
\label{zschdef}
\eea
where $\Lambda$ is an arbitrary fixed scale, and the subtraction is
necessary in order to ensure that $\ln
Z^{\overline{\rm res}}$ be free of linear $\ln N$ terms, in agreement
with Eq.~(\ref{zterm}).

Resummed anomalous dimension and coefficient function in any scheme
can in general be obtained from the full physical scheme anomalous
dimension, 
which combines
resummed and unresummed results (such as Eq.(\ref{nlopadres})) by
performing the combination of a resummed (\ref{zresdef})
and unresummed (\ref{zexp}) scheme change,
\beq
Z(N,\as)=Z^{\rm u}(N,\as) Z^{\rm res}(N,\as).
\label{genschch}
\eeq 
In practice, in the sequel we will take $Z^{\rm res}$ to be of the form
Eq.~(\ref{zschdef}). Then, $Z^{\rm res}$
contains higher order powers of $\ln N$ but no linear terms in
$\ln N$ (nor, of course,  non-logarithmic terms). 
Furthermore, we will take $Z^{\rm u}$ to 
contain at most a linear term in $\ln N$, but not higher
order powers. In such case, there is no overlap
between a resummed next$^k$-to-leading log scheme change $Z^{\rm res}$
and an unresummed next$^k$-to-leading order scheme
change  $Z^{\rm u}$:
the unresummed and resummed scheme
changes affect different contributions in an expansion of the
anomalous dimension in powers of $\as$ and $\ln N$.

Let us now consider explicitly the  commonly used \MS\ scheme.
This scheme  is
characterized by the fact that $Z^{(C)}$ Eq.~(\ref{mulnor}) is a
series of pure poles in $1/\bar\epsilon\equiv
1/\epsilon-\gamma_E+\log(4\pi)$.  It can then be shown~\cite{kor,albal}
that if a resummation of the form of Ref.~\cite{contopa} holds (\ie,
of the form Eq.~(\ref{finresad}) with all $g_{ij}=0$ when $i>0$), the
\MS\ anomalous dimension $\gamma^{\rm AP}$ has, to all orders in
$\as$, at most a single $\ln(1/N^a)$. In order to get from the physical
to the \MS\ scheme  one has to perform a suitable combination of 
the well--known unresummed
scheme change 
which takes to \MS\ from the so--called DIS scheme~\cite{dflm}, and
a resummed scheme change. This latter
resummed 
scheme change is in fact entirely determined by the requirement that
the \MS\ anomalous dimension be 
linear in $\ln(1/N^a)$.

To see
this, assume that the resummed anomalous dimension in the physical
scheme is given by
\beq
\gamma^{\rm AP,res}(N,\as(\mu^2))=\int_1^{N^a}\frac{dn}{n}\,
\sum_{j=1}^\infty g_{0j}\, \as^j(\mu^2/n),
\label{gamresphy}
\eeq
and perform a resummed  scheme change
Eqs.~(\ref{zrresdef},\ref{zschdef}), with $z_{ij}=0$ for $i>0$.
After such a scheme change,
\bea
{{\gamma}^\prime}^{\rm AP,res}
(N,\as(\mu^2))&=&\int_1^{N^a}\frac{dn}{n}\,
\sum_{j=1}^\infty g_{0j}\, \as^j(\mu^2/n)
+\int_1^{N^a} \frac{dn}{n}\,\sum_{j=1}^\infty z_{0j} 
\left[\as^j(\mu^2/n)-\as^j(\mu^2)\right]
\nonumber\\
&=&\int_1^{N^a}\frac{dn}{n}\,
\sum_{j=1}^\infty \left(g_{0j}+z_{0j}\right) \as^j(\mu^2/n)
+\ln\frac{1}{N^a}\,\sum_{j=1}^\infty z_{0j} \as^j(\mu^2).
\eea
Hence, imposing that
${\gamma^\prime}^{\rm AP,res}$  be
linear in $\ln(1/N^a)$ fixes
\beq
Z^{\rm res}=-\sum_{j=1}^\infty g_{0j} \as^j(k^2/n).
\label{msschch}
\eeq

Performing the scheme change Eq.~(\ref{msschch}) one gets
\bea
{{\gamma}^\prime}^{\rm AP,res}
(N,\as(\mu^2))
&=&-\ln\frac{1}{N^a} \sum_{j=1}^\infty g_{0j}\, \as^j(\mu^2),
\label{gamresms}\\
{{C}^\prime}^{\rm res}\left(N,1,\as(Q^2)\right)
&=&
%\exp\sum_{j=1}^\infty g_{0j}\int_1^{N^a}\frac{dn}{n}\,
%\left[\int_{\mu^2}^{Q^2}\frac{dk^2}{k^2}\as^j(k^2/n)
%+\int_{\mu^2 n}^{\mu^2}\frac{dk^2}{k^2}  \as^j(k^2/n)\right]
%\nonumber\\
\exp\sum_{j=1}^\infty g_{0j}\int_1^{N^a}\frac{dn}{n}\,
\int_{Q^2n}^{Q^2}\frac{dk^2}{k^2}\as^j(k^2/n).
\label{cresms}
\eea
The resummed \MS\ anomalous dimension and coefficient function can now 
be obtained  by transforming those given in
Eqs.~(\ref{gamresms},\ref{cresms})  with 
an unresummed scheme change obtained by using $Z^{\rm
u}_{\rm \overline{MS}}$ of
Ref.~\cite{dflm}, but
omitting all contributions to
$Z^{\rm
u}_{\rm \overline{MS}}$ which are proportional to
$\ln^m(1/N^a)$ with $m>1$, so that the requirement on $Z^{\rm
u}$ discussed after Eq.~(\ref{genschch}) is satisfied. 

In practice, however, it is easier to observe that the requirement
that the \MS\ anomalous dimension be free of higher order powers of
$\ln N$ simply means that
the \MS\ resummed anomalous dimension coincides with the
unresummed one. This further implies that 
the \MS\ resummed coefficient function is found
combining the resummed result Eq.~(\ref{cresms}) with the standard
\MS\ 
unresummed
coefficient function, and subtracting double counting. 
At next-to-leading order, next-to-leading log we get
\bea
C_{\overline{\rm MS}}\left(N,1,\as(Q^2)\right)&=&
\exp\left[\int_1^{N^a} \frac{dn}{n}
\int_{Q^2 n}^{Q^2}\frac{dk^2}{k^2} \left(
g_{01} \as(k^2/n)+g_{02} \as^2(k^2/n)\right)\right]\nonumber\\
%&&\phantom{aaaaaaaaaaa}
&\times&\left[1+ \as(Q^2)
\left(C_1(N,1)+\frac{g_{01}}{2} \ln^2\frac{1}{N^a}\right)\right]
+O(\as^{k+2}\ln^k N)+O(\as^2)
\nonumber\\
\phantom{C_{\overline{\rm MS}}\left(N,Q^2/\mu^2,\as(\mu^2)\right)}
&=&\exp\left[\int_1^{N^a} \frac{dn}{n}
\int_{Q^2 n}^{Q^2}\frac{dk^2}{k^2} \left(
g_{01} \as(k^2/n)+g_{02} \as^2(k^2/n)\right)\right]
\nonumber\\
%&&\phantom{aaaaaaaaaaa}
&+&\as(Q^2) \left(C_1(N)+\frac{g_{01}}{2} \ln^2\frac{1}{N^a}\right)
+O(\as^{k+2}\ln^k N)+O(\as^2)
\label{cmsbarfinb}
\eea
where $C_1(N)$ is the standard next-to-leading order contribution to
the coefficient function. In the first line on the r.h.s.
we have given the
result which is obtained by performing the two scheme changes $Z^{\rm
res}_{\rm \overline{MS}} Z^{\rm u}_{\rm \overline{MS}}$, which
correspond respectively to the first and second factor in square
brackets, 
and in the second line the result
obtained by combining Eq.~(\ref{cresms}) with the unresummed result
and explicitly performing the double-counting subtraction; of course
the two results agree up to subleading corrections.

Resummed results in the \MS\ scheme can be analogously obtained
starting from the expression Eq.~(\ref{finresspf}) of the resummed
physical anomalous dimension in terms of a resummed splitting
function. Proceeding in a similar way, with obvious modifications,
we get
\bea
&&C_{\overline{\rm MS}}\left(N,1,\as(Q^2)\right)=
\exp\left[a\int_0^1 \! dx \frac{x^{N-1}-1}{1-x}
\int_{Q^2}^{Q^2(1-x)^a}\frac{dk^2}{k^2} \left(
\hat g_{01} \as(k^2)+\hat g_{02}
\as^2(k^2)\right)\right]\nonumber\\
&&\qquad
+ \as(Q^2)
\int_0^1 x^{N-1} \left[C_1(x)
-\frac{\hat g_{01}}{2} \left(\frac{\ln(1-x)^a}{1-x}\right)_+\right]
+O(\as^{k+2}\ln^k N)+O(\as^2),
\label{cmsbarfinx}
\eea
where now $\hat g_{0i}$ are contributions
to the coefficient of the
$O(\ln(1-x)^a)$ term in the physical  splitting function, and
$C_1(x)$ is the $x$-space next-to-leading order contribution to
the coefficient function. Note that the double-counting subtraction
in Eq.~(\ref{cmsbarfinb}) is not just the Mellin transform of the
double counting subtraction of Eq.~(\ref{cmsbarfinx}): this is only
true at the leading log level.
Again, the two forms Eq.~(\ref{cmsbarfinb}) and~(\ref{cmsbarfinx}) of
the resummed \MS\ coefficient function coincide at the next-to-leading
order and next-to-leading log level, but differ by terms which are at
least of $O(\as^3)$ and then either next-to-next-to-leading log or
suppressed by powers of $N$. 

The resummed result Eq.~(\ref{cmsbarfinx}) can be finally brought in the form
of Ref.~\cite{cnt} by recalling that $\hat g_{02}$ is determined in
terms of the physical anomalous dimension, and
separating off the contributions which originate from
the anomalous dimension and coefficient function, according to
Eq.~(\ref{gAB}):
\bea
&&C_{\overline{\rm MS}}\left(N,1,\as(Q^2)\right)
\label{cmsbarcnt}
\\&&=
\exp a\int_0^1 \! dx \frac{x^{N-1}-1}{1-x}
\left[\int_{Q^2}^{Q^2(1-x)^a}\frac{dk^2}{k^2} 
\left[A_1 \as(k^2)+ A_2 \as^2(k^2)\right]+ B_1 \as(Q^2(1-x)^a)
\right]
\nonumber\\
&&+\as(Q^2)
\int_0^1 x^{N-1} \left[C_1(x)-  
\frac{A_1}{2} \left(\frac{\ln(1-x)^a}{1-x}\right)_+ 
-\left(\frac{B_1}{1-x}\right)_+\right] 
+O(\as^{k+2}\ln^k N)+O(\as^2).
\nonumber
\eea
Note that the double counting subtraction is not the same as in
Eq.~(\ref{cmsbarfinx}), 
because the $B$-term in
Eq.~(\ref{cmsbarcnt}) is not free of $O(\ln\frac{1}{N})$ terms.

The physical and \MS\ schemes which we discussed explicitly are
extreme choices in the
treatment of the resummed terms: in the former case, these are entirely
included in the anomalous dimension, and in the latter case they are
entirely included in the coefficient function. Of course, a variety of
intermediate choices is possible. These resummed scheme 
choices can then be freely
combined with the usual choice of unresummed factorization scheme.
For instance, one
can combine an \MS-like resummation (where the resummation is
included in the coefficient function) with a physical-scheme anomalous
dimension. In such case, the anomalous dimension is given, up to
$O(N^0)$ terms, by Eq.~(\ref{gamresms}). Clearly, all such scheme choices can
be obtained by a combination of the resummed and unresummed scheme
changes Eqs.~(\ref{zexp},\ref{zterm}).

\sect{Summary}

In this paper, we have discussed the resummation of large logarithmic
contributions to perturbative coefficients that arise as a left-over
of the cancellation of infrared singularities near the boundary of the
phase space. Our approach is based on identifying the relevant
dimensionful scale for the large logs,
showing that it is  determined essentially by kinematic
considerations, and using the renormalization group to resum it.

The main result of this paper is  a resummation formula which allows
the computation of the physical anomalous dimension
(\ie, the scale dependence of physical observables)
to any logarithmic
accuracy, starting from a fixed-order calculation. This result is less
predictive than what one gets~\cite{contopa} by  extending the structure
of familiar next-to-leading log resummation~\cite{cnt,sterman} 
to high orders:
specifically, we find that   an order $\frac{k(k+1)}{2}$ 
perturbative calculation is
required for resummation to next$^{k-1}$-to-leading log accuracy. In
particular, the standard next-to-leading log resummation turns out to
be fully determined by the fixed next-to-leading order result only
because of the vanishing of a coefficient which, in our approach,
appears to be
accidental. 

The advantage of this approach, however, is that it does not rely on
factorization properties of the physical cross section. Indeed, our
proof does not require any detailed consideration of the individual
Feynman diagrams which contribute in the soft limit,  and thus in
particular  it does not
require the classification of the different forms of radiation
which contribute in the soft limit (soft radiation, 'jet' radiation
collinear to incoming or outgoing partons etc.) which is the main
complication in conventional next-to-leading~\cite{cnt,sterman} or
all-order~\cite{contopa} approaches to resummation. 

An  important feature of the present approach
is that resummed results are  provided
directly for a physical observable, the physical anomalous
dimension, rather than for partonic quantities. 
This makes it easier to study issues related to the choice of
an  appropriate
resummation variable. In particular, we have shown that a leading (or
next-to-leading, etc.)  $\ln(1-x)$ expression of the cross section is
necessarily ill-defined, and traced the origin of this problem, first
pointed out in Ref.~\cite{paolo}, to subleading $\ln(1-x)$ terms which
appear in the inverse Mellin transform of a leading $\ln N$
expression, which is instead well-defined.
Also, the discussion of issues related to the choice of
factorization scheme is considerably streamlined: our results
are manifestly scheme-independent, and 
results within specific choices of 
scheme can be readily obtained  as special cases. 

Because of its generality, 
the approach to soft resummation presented in this paper
lends itself naturally to a variety
of extensions and generalizations.  In particular, it should be easy
to generalize it to other classes of soft resummation, such as 
jet or prompt-photon production, and, perhaps more interestingly, to the
resummation of transverse momentum distributions, which can be
unified~\cite{contopa} 
with the resummation discussed here by considering
Mellin and Fourier transforms as special cases of a complex integral transform.

It is interesting to ask whether the formalism presented in this paper
might also lead to
further insight on resummation at small $x$.

\section*{Acknowledgements}
We thank G.~Altarelli, R.~Ball, C.~Becchi, L.~Magnea, J.~I.~Latorre
and L.~Trentadue for various
discussion during the course of this work. We are especially grateful
to S.~Catani for many stimulating observations and criticism of a draft
of the paper.

\vfill\eject
\appendix
\section{N-body phase space}
\label{phsp}
The $n$-body phase space can be expressed in terms of the $n-1$-body
and  the two-body phase
spaces~\cite{gambnotes}. 
To see this, start with the standard definition of the phase space for
a generic process 
with incoming momentum $P$ and $n$ bodies  in the final state 
with outgoing momenta $k_1$,\dots,$k_n$ and masses $k^2_i=m^2_i$:
\beq
d\phi_{n}(P;k_1,\ldots,k_n)
=\frac{d^{d-1}k_1}{(2\pi)^{d-1}2k_1^0}\ldots
\frac{d^{d-1}k_n}{(2\pi)^{d-1}2k_n^0}
(2\pi)^d\,\delta^{(d)}(P-k_1-\ldots-k_n),
\label{npsp}
\eeq

The momentum-conservation delta function can be rewritten as
\beq
\delta^{(d)}(P-k_1-\ldots-k_n)
=\int d^d P_n\,\delta^{(d)}(P-P_{n-1}-k_n)\,
\delta^{(d)}(P_{n-1}-k_1-\ldots-k_{n-1}).
\label{momdel}
\eeq
Now note that
\beq
d^dP_{n-1} = (2\pi)^d\,\frac{d^{d-1}P_{n-1}}{(2\pi)^{d-1}2P^0_{n-1}}\,
\frac{d(P^0_{n-1})^2}{2\pi}
 = (2\pi)^d\,\frac{d^{d-1}P_{n-1}}{(2\pi)^{d-1}2P^0_{n-1}}\,
\frac{dM_{n-1}^2}{2\pi},
\label{dpn}
\eeq
where
\beq
M_{n-1}^2 = (P_{n-1}^0)^2 - |\vec{P}_{n-1}|^2;\qquad
\left(m_1+m_2+\dots+m_{n-1}\right)^2
\leq M_{n-1}^2\leq\left(\sqrt{P^2}-m_n\right)^2.
\label{mdeflim}
\eeq
We obtain immediately
\beq
d\phi_{n}(P;k_1,\ldots,k_n)=\int_0^s\frac{dM^2_{n-1}}{(2\pi)}\,
d\phi_2(P;P_{n-1},k_n)\,\,d\phi_{n-1}(P_{n-1};k_1,\ldots,k_{n-1}),
\label{ntonmo}
\eeq
which is the desired result. Using this result recursively, the $n$
body phase space can be entirely expressed in terms of two-body phase
space integrals.

For completeness, we also give the expression of the two-body phase
space in the center-of-mass frame. We have
\bea
d\phi_2(P_{i+1};P_i,k_i)&=&
\frac{d^{d-1}P_i}{(2\pi)^{d-1}\,2P^0_i}\,
\frac{d^{d-1}k_i}{(2\pi)^{d-1}\,2k_i^0}\,
(2\pi)^d \delta^{(d)}(P_{i+1}-P_i-k_i)
\nonumber\\
&=&\frac{(2\pi)^{2\epsilon-2}}{4}\,
\frac{d^{d-1}k_i}{P^0_i k^0_i}\,
\delta(P^0_{i+1}-P^0_i-k^0_i).
\label{tbps}
\eea
In the center-of-mass frame, $\vec{P}_{i+1}=0$, $P^0_{i+1}=M_{i+1}$ and 
$|\vec{P}_i|=|\vec{k}_i|=k_i^0$, since $k_i^2=0$.
We have therefore
\beq
\delta(P^0_{i+1}-P^0_i-k^0_i)=\frac{\sqrt{(k^0_i)^2+M_i^2}}{M_{i+1}}\,
\delta\left(k_i^0-\frac{M_{i+1}^2-M_i^2}{2M_{i+1}}\right).
\eeq
Hence
\beq
\label{kiz}
|\vec{k}_i|=k_i^0=\frac{M_{i+1}^2-M_i^2}{2M_{i+1}}
=\frac{M_{i+1}}{2}\left(1-\frac{M_i^2}{M^2_{i+1}}\right),
\eeq
and
\beq
d\phi_2(P_{i+1};P_i,k_i)=
\frac{(2\pi)^{2\epsilon-2}}{4}\,
\frac{(k_i^0)^{1-2\epsilon}}{M_{i+1}}\,d\Omega_i
=
\frac{1}{2(4\pi)^{2-2\epsilon}}\,
M_{i+1}^{-2\epsilon}\,
\left(1-\frac{M_i^2}{M^2_{i+1}}\right)^{1-2\epsilon}\,d\Omega_i.
\label{ftbs}
\eeq
  
\eject

\begin{thebibliography}{99}  
  
\baselineskip14pt  
  
\bibitem{sudakov} 
V.~V.~Sudakov,
Sov.\ Phys.\ JETP {\bf 3} (1956) 65.
%%CITATION = SPHJA,3,65;%%
\bibitem{resrev} 
See \eg\ 
S.~Catani {\it et al.},
%%CITATION = HEP-PH 0005025;%%
{\tt hep-ph/0005025} and references therein.
\bibitem{loexp}
G.~Parisi,
%``Summing Large Perturbative Corrections In QCD,''
Phys.\ Lett.\ B {\bf 90} (1980) 295;\\
%%CITATION = PHLTA,B90,295;%%
G.~Curci and M.~Greco,
%``Large Infrared Corrections In QCD Processes,''
Phys.\ Lett.\ B {\bf 92} (1980) 175.
%%CITATION = PHLTA,B92,175;%%
\bibitem{cnt} 
S.~Catani and L.~Trentadue,
Nucl.\ Phys.\ B {\bf 327} (1989) 323.
%%CITATION = NUPHA,B327,323;%%
\bibitem{sterman}
G.~Sterman,
Nucl.\ Phys.\ B {\bf 281} (1987) 310.
%%CITATION = NUPHA,B281,310;%%
\bibitem{equiv}
S.~Catani and L.~Trentadue,
Nucl.\ Phys.\ B {\bf 353} (1991) 183.
%%CITATION = NUPHA,B353,183;%%
\bibitem{paolo}
S.~Catani, M.~L.~Mangano, P.~Nason and L.~Trentadue,
Nucl.\ Phys.\ B {\bf 478} (1996) 273.
%%CITATION = HEP-PH 9604351;%%     
\bibitem{contopa}
H.~Contopanagos, E.~Laenen and G.~Sterman,
Nucl.\ Phys.\ B {\bf 484} (1997) 303.
%%CITATION = HEP-PH 9604313;%%
\bibitem{fact}
J.~C.~Collins, D.~E.~Soper and G.~Sterman,
Adv.\ Ser.\ Direct.\ High Energy Phys.\  {\bf 5}~(1988)~1.
%%CITATION = 00319,5,1;%%
\bibitem{physcat}
S.~Catani,
%``Physical anomalous dimensions at small x,''
Z.\ Phys.\ C {\bf 75} (1997) 665.
%%CITATION = HEP-PH 9609263;%%
%\cite{Lee:is}
\bibitem{ircan}
T.~Kinoshita,
%``Mass Singularities Of Feynman Amplitudes,''
J.\ Math.\ Phys.\  {\bf 3} (1962) 650;\\
T.~D.~Lee and M.~Nauenberg,
%``Degenerate Systems And Mass Singularities,''
Phys.\ Rev.\  {\bf 133} (1964) B1549.
%%CITATION = PHRVA,133,B1549;%%
%%CITATION = JMAPA,3,650;%%
%\bibitem{f2cf} 
%G.~Altarelli, R.~K.~Ellis and G.~Martinelli,
%``Leptoproduction And Drell-Yan Processes Beyond 
% The Leading Approximation In Chromodynamics,''
%Nucl.\ Phys.\ B {\bf 143} (1978) 521
%[Erratum-ibid.\ B {\bf 146} (1978) 544].
\bibitem{IZ}
See \eg\ C.~Itzykson and J.~B.~Zuber, ``Quantum Field Theory''
(McGraw-Hill, New York, 1980) Sect.~6.2.3, and Ref. therein.
\bibitem{dynnlo}
T.~Matsuura, S.~C.~van der Marck and W.~L.~van Neerven,
%``The Calculation Of The Second Order Soft And Virtual Contributions To The Drell-Yan Cross-Section,''
Nucl.\ Phys.\ B {\bf 319} (1989) 570;\\
%%CITATION = NUPHA,B319,570;%%
W.~L.~van Neerven and E.~B.~Zijlstra,
%``The O (alpha-s**2) corrected Drell-Yan K factor in the DIS and MS scheme,''
Nucl.\ Phys.\ B {\bf 382} (1992) 11.
%%CITATION = NUPHA,B382,11;%%
\bibitem{disnnlo}
E.~B.~Zijlstra and W.~L.~van Neerven,
%``Order alpha-s**2 QCD corrections to the deep inelastic proton structure functions F2 and F(L),''
Nucl.\ Phys.\ B {\bf 383} (1992) 525.
%%CITATION = NUPHA,B383,525;%%
\bibitem{cfp}
E.~G.~Floratos, D.~A.~Ross and C.~T.~Sachrajda,
%``Higher Order Effects In Asymptotically Free Gauge Theories: The Anomalous Dimensions Of Wilson Operators,''
Nucl.\ Phys.\ B {\bf 129} (1977) 66
[Erratum-ibid.\ B {\bf 139} (1978) 545];\\
%%CITATION = NUPHA,B129,66;%%
G.~Curci, W.~Furmanski and R.~Petronzio,
%``Evolution Of Parton Densities Beyond Leading Order: The Nonsinglet Case,''
Nucl.\ Phys.\ B {\bf 175} (1980) 27;\\
%%CITATION = NUPHA,B175,27;%%
E.~G.~Floratos, C.~Kounnas and R.~Lacaze,
%``Higher Order QCD Effects In Inclusive Annihilation And Deep Inelastic Scattering,''
Nucl.\ Phys.\ B {\bf 192} (1981) 417.
%%CITATION = NUPHA,B192,417;%%
\bibitem{kor}
G.~P.~Korchemsky,
Mod.\ Phys.\ Lett.\ A {\bf 4} (1989) 1257.
%CITATION = MPLAE,A4,1257;%%
\bibitem{scia} C.~F.~Berger, {\tt hep-ph/0209107};\\
%%CITATION = HEP-PH 0209107;%%
S.~Moch, J.~A.~Vermaseren and A.~Vogt,
%``Non-Singlet Structure Functions at Three Loops: Fermionic Contributions,''
{\tt hep-ph/0209100}.
%%CITATION = HEP-PH 0209100;%%
\bibitem{dyres}
L.~Magnea,
Nucl.\ Phys.\ B {\bf 349} (1991) 703.
%%CITATION = NUPHA,B349,703;%%
\bibitem{albal}
S.~Albino and R.~D.~Ball,
Phys.\ Lett.\ B {\bf 513} (2001) 93.
%%CITATION = HEP-PH 0011133;%%
\bibitem{dflm}
M.~Diemoz, F.~Ferroni, E.~Longo and G.~Martinelli,
Z.\ Phys.\ C {\bf 39} (1988) 21.
%%CITATION = ZEPYA,C39,21;%%
\bibitem{gambnotes}
G.~Gamberini, {\it unpublished notes}.

\end{thebibliography}
\end{document}